%% file: main.tex
\documentclass[journal]{IEEEtran}
 \usepackage{amsfonts,epsfig,multirow,amssymb,amsmath}
 \usepackage[normalem]{ulem}
\usepackage{bbm,url}
\usepackage{framed}
\usepackage{algcompatible}
\usepackage{algpseudocode}
\usepackage[]{xcolor}
\usepackage{graphicx}
\usepackage{lineno}   
\usepackage[normalem]{ulem} 
\usepackage{dsfont}
\usepackage{epstopdf}  
\usepackage{amsthm}
\usepackage{color}
\usepackage{flushend}
\usepackage{amsmath,amssymb,latexsym,amsthm}
\usepackage{mathtools}
\usepackage{comment}

\usepackage{bbm}
\usepackage{dsfont}
\usepackage[normalem]{ulem}
\useunder{\uline}{\ul}{}

\usepackage[linesnumbered,ruled,vlined]{algorithm2e}

\SetKwInput{KwInput}{Input}                % Set the Input
\SetKwInput{KwOutput}{Output}              % set the Output

\newtheorem{lemma}{Lemma}

\graphicspath{{FinalFigures/}}

\begin{document}

\title{Cost-aware Feature Selection for\\  IoT Device Classification}

%\title{Feature Selection for IoT Device Classification via Risk Minimization under Budget Constraint}

%\author{Biswadeep Chakraborty$^{1}$, Dinil Mon Divakaran$^{2}$, Ido Nevat$^{3}$, Gareth W. Peters$^{4}$ and Mohan Gurusamy$^{1}$\\
%$^{1}$ National University of Singapore,
%$^{2}$ Trustwave,
%$^{3}$ TUMCREATE (Singapore)\\
%$^{4}$ Department of Actuarial Mathematics and Statistics, Heriot-Watt University (United Kingdom)}

\author{Biswadeep Chakraborty, Dinil Mon Divakaran, Ido Nevat, Gareth W. Peters, Mohan Gurusamy
\thanks{Biswadeep Chakraborty (bchakraborty6@gatech.edu) is with Georgia Institute of Technology.  Dinil Mon Divakaran  (dinil.divakaran@trustwave.com; corresponding author) is with Trustwave (Singapore). Ido Nevat (ido.nevat@tum-create.edu.sg) is with TUMCREATE (Singapore). Gareth W. Peters (g.peters@hw.ac.uk) is with Department of Actuarial Mathematics and Statistics, Heriot-Watt University (United Kingdom). Mohan Gurusamy (gmohan@nus.edu.sg) is with National University of Singapore (Singapore).}
}

\markboth{The paper is accepted for publication at IEEE IoT Journal.}%      
{Chakraborty \MakeLowercase{\textit{et al.}}}  
\maketitle

\begin{abstract}
\input{tex/abstract}
\end{abstract}

% \begin{IEEEkeywords}
% Optimization, IoT, Identification, Network
% \end{IEEEkeywords}

%%%%%%%%%%%%%%%%%%%%%%%%%%%%%%%%%%%%%%%%%%%%%%%%%%%%%%%%%%%%%%%%%%%%%%%%%%%%%%%%%%%%%%%%%%%%%%

%\thispagestyle{empty}
\pagestyle{empty}
\pagestyle{plain}

%%%%%%%%%%%%%%%%%%%%%%%%%%%%%%%%%%%%%%%%%%%%%%%%%%%%%%%%%%%%%%%%%%%%%%%%%%%%%%%%

\input{tex/intro}

\input{tex/related}

\input{tex/system}

\input{tex/algos}

\input{tex/perf-eval}

\input{tex/conclusions}

%%%%%%%%%%%%%%%%%%%%%%%%%%%%%%%%%%%%%%%%%%%%%%%%%%%%%%%%%%%%%%%%%%%%%%%%%%%%%%%%

\vspace{-0.2cm}
\section*{Acknowledgement}
This research is supported by the National Research Foundation, Prime Minister's Office, Singapore under its Corporate Laboratory@University Scheme, National University of Singapore, and Singapore Telecommunications Ltd.

\bibliographystyle{IEEEtran}
\bibliography{references}

\input{tex/appendix}

\end{document}

%% file: tex/abstract.tex
Classification of IoT devices into different types is of paramount importance, from multiple  perspectives, including security and privacy aspects. 
%IoT devices easy targets for large-scale compromises for malicious activities such as privacy breach, data exfiltration, bot infection, attack launches, etc. 
Recent works have explored machine learning techniques for fingerprinting (or classifying) IoT devices, with promising results. However, existing works have assumed that the features used for building the machine learning models are readily available or can be easily extracted from the network traffic; in other words, they do not consider the costs associated with feature extraction. In this work, we take a more realistic approach, and argue that feature extraction has a cost, and the costs are different for different features. 
We also take a step forward from the current practice of considering the misclassification loss as a binary value, and make a case for different losses based on the misclassification performance. Thereby, and more importantly, we introduce the notion of \textit{risk} for IoT device classification. We define and formulate the problem of cost-aware IoT device classification. This being a combinatorial optimization problem, we develop a novel algorithm to solve it in a fast and effective way using the Cross-Entropy (CE) based stochastic optimization technique. Using traffic of real devices, we demonstrate the capability of the CE based algorithm in selecting features with minimal risk of misclassification while keeping the cost for feature extraction within a specified limit. 

%% file: tex/intro.tex
\section{Introduction}

The rapid growth of the Internet of Things (IoT) market is also having an impact on the threat landscape of the cyber space~\cite{symantec-iot-attack-report-2019}. 
We now have attacks that compromise large numbers of IoT devices, subsequently using them as bots for launching different kinds of large-scale attacks (e.g., DDoS attacks) that result in significant losses~\cite{Understanding-Mira-2017, hajime-NDSS-2019}. Attackers often explore and exploit vulnerabilities to gain access to machines. In this IoT era, it is now possible to exploit a few vulnerabilities of just one device type, and compromise hundreds of thousands of devices of the same type using the same exploits.

To detect and prevent attacks on and from IoT devices, one of the important tasks is to identify the type of devices connected to a network. 
%(of, say an enterprise). 
Device identification is useful in keeping track of various device types in a network, which consequently helps in analyzing and defending against potential vulnerabilities of various IoT devices~\cite{NDSS-DISS-ADROIT-2020}.
%, NDSS-DISS-BlockChain-2020}.
In addition, a mitigation solution---such as blocking devices with newly known vulnerabilities---can be implemented quickly, if the network administrator knows the identities (brand, model number, functionality, etc.) of the devices connected to the network. Recent years have seen research proposals on identifying IoT devices, particularly by analyzing their network traffic (e.g., see~\cite{Meidan2017ProfilIoT,miettinen2017iot,fingeprinting-behind-NAT-VPN-2019} and \cite{audi-JSAC-2019}). 
Existing solutions often use machine-learning based approaches to fingerprint IoT devices, by extracting a number of {\em features} from IoT traffic. For instance, in the case of supervised or semi-supervised approaches (e.g.,~\cite{dev-classification-Trans-MC-2018} and~\cite{DEFT-2019}), the features extracted are used to train a classification model to differentiate IoT devices while in operation. 
%Features from network traffic are also used for machine learning based anomaly detection in IoT networks~\cite{IoT-keeper-2018, DIoT-ICDCS-2019}.

A common underlying assumption of the device classification works in the literature is that all features can be easily extracted from the traffic and that there is no difference---technical or otherwise---in extracting any two different features. However, we argue that, such an assumption misses out some important points. Extraction of features from traffic incurs a cost, and different features have different costs for extraction. We identify three types of feature-extraction costs: \\

\noindent 1. {\em Computational cost:} When network traffic characteristics are used for classifying devices, the relevant features are computed by reading and processing traffic at a router, a gateway or a server to which the traffic is mirrored to. That is, there is cost involved in computing a feature from packets, and different features may incur different computational costs. For example, searching for a specific pattern, say, malware signature/hash, in packets is computationally more expensive than counting the number of packets. 
    
\noindent 2. {\em Memory cost:} Feature extraction also involves use of memory to store the (running) value of a feature. For example, storing the number of packets traversing a link requires just one counter, but to count the number of packets of each connection that is active at a link requires maintenance of a hash table or some variants~\cite{HashPipe-SOSR-2017}. And memory is an extremely valuable resource in routers and gateways. Indeed, traditionally ISP routers use sampled NetFlow~\cite{RFC7011} to capture and aggregate meta-information of traffic flows,  due to resource constraints at the router. With NetFlow, information which is often assumed to be easily available (e.g., size of each packet in a connection) is not collected. 
%Now, with programmable data planes, there are renewed efforts to come up with new data structures and algorithms for tracking connection level information (such as heavy hitters)~\cite{NitroSketch-SIGCOMM-2019}. 
    
\noindent 3. {\em Privacy cost:} The third factor that decides the cost of a feature is privacy. Some features, such as those extracted from payloads of (unencrypted) network traffic, can reveal sensitive or even confidential information of end-users~\cite{smart-home-privacy-2016}. In fact, DNS or even DNS over HTTPS (that is, DNS over encrypted HTTP connection) can reveal sensitive information related to users communication~\cite{DNS-ads-profit-2011, DoH-2019}. Privacy concerns imply that, either some features might be unavailable (i.e., cost is infinite) or there would be some cost incurred to obtain such data, for example by paying for the data with user's consent or by anonymizing the data to minimize privacy leak. %Anomymization is an operation that has cost associated with it; but it also results in lossy information. 

%As per recent policies \todoalso results in lossy information. {cite GDPR}, there is restriction in storing data that identifies with consumers. 

Therefore, we argue that, a solution for device identification (or classification) should take the cost of features into consideration. This in turn means, the performance of a device classification solution is  dependent on the cost of features used, since the cost may exclude some important features. 
A solution developer would need to be conscious of the budget available for deploying the solution. 

Previous works also assume no difference in device misclassification, in that they were not concerned by the misclassification class to which an incorrect assignment was made. In practice, not only the fact that a misclassification took place is important, it is also important to consider which class the device was misclassified under as this could have ramifications for actions and costs. Previous works also assume no difference in device misclassification; i.e., they were not concerned of the type a given device is being misclassified into. However, such an approach overlooks important aspects IoT devices. An IoT device, say a camera, might be from vendor X with model name Y; but the same vendor might have multiple models for the product camera. Consider the different misclassifications possible: (i)~[camera, X, Y] gets classified as smart bulb, (ii)~[camera, X, Y] gets classified as [camera, A, B], and (iii)~[camera, X, Y] gets classified as [camera, X, Z]. Clearly, the third misclassification is more acceptable than the first two, since the functionality and brand name were correct; and the first misclassification is the worst result to have among the three. Therefore, we argue the need to consider {\em multiple losses} due to misclassifications that essentially captures the differences in misclassifications.

In this work, we address the problem of classifying IoT devices, under the realistic assumptions that (i)~features have associated costs, (ii)~a solution developer has a budget constraint, and (iii)~there could be different kinds of losses due misclassifications. Different from existing works, the challenge here is to develop a cost-optimal IoT device classification solution with high accuracy. 

Our contributions are the following:
%\todo{@Biswadeep: list down}
\begin{enumerate}
    \item We introduce the notion of feature cost and budget constraint in the problem of IoT device classification. Furthermore, going beyond the current way of treating misclassification as a binary value, we consider multiple losses due to misclassification (Sec.~\ref{subsec:loss}), and introduce the notion of {\em risk} (Sec.~\ref{subsec:risk}) to evaluate a classifier's performance. Subsequently, we formulate the feature-selection problem under constrained budget as an optimization problem (Section~\ref{sec:problem}). 
    
    \item We present and develop a cross-entropy (CE) based algorithm (Section~\ref{sec:cross}), that solves the optimization problem efficiently. In comparison to a brute force approach which has to run a classifier exponential number of times, the number of executions (of classifier) that the CE algorithm has to run is linear in its two parameters. 

    \item We conduct extensive experiments using traffic of real IoT devices, analyzing a brute force approach, multiple greedy algorithms and our proposed CE-based algorithm, comprehensively. 
\end{enumerate}

We believe our framework for cost-optimal feature selection is also applicable for other problems, where features have associated costs and solution deployment has a budget constraint. Examples of such problems in the domain of network traffic analysis include the traditional network application classification~\cite{SLIC-2015}, botnet detection~\cite{disclosure-ACSAC-2012}, anomaly detection in traditional~\cite{NADA-2018, GEE-CNS-2019} as well as IoT~\cite{DIoT-ICDCS-2019} networks, etc. 
%\todo{incomplete}

%\todo{paper organization}
%In Section \ref{Sec:related} we discuss related works on IoT device classification. 
After defining the system model in Section~\ref{sec:system},   we formulate the optimisation problem in Section \ref{sec:problem}. In Section~\ref{sec:cross}, we develop the CE-based algorithm. Subsequently, in Section~\ref{sec:others}, we present three greedy algorithms for selecting features while respecting the provided budget. We carry out experiments and present results in Section~\ref{sec:simulations}.

%% file: tex/related.tex
\section{Related works}\label{Sec:related}

Fingerprinting or identification of hosts, operating systems, etc. has been a problem of interest for many years now (e.g., see~\cite{shamsi2014hershel}). Similarly, there are works on fingerprinting devices; for instance 
Gao {\em et al.}~\cite{gao2010passive} proposed to identify wireless access points by applying a wavelet-based approach on frame arrival time differences. With the emergence of IoT devices, research works have looked into inferring different aspects of IoT devices from their network traffic~\cite{maiti2017link, apthorpe2017spying}. For example, the work in~\cite{apthorpe2017spying} reveals private and sensitive information of users at a smart home are leaked by analyzing only the encrypted network traffic. 

Identification of IoT devices is of use to different entities, such as enterprises and ISPs. For an enterprise, identification helps in asset tracking as well as for securing the devices from potential vulnerability exploitation; whereas, for ISPs, knowing the type of different devices connected to its network might help in mitigation of large-scale attacks. For instance, Yang {\em et al.}~\cite{yang2019towards} generated fingerprints of devices using neural algorithms, which were used to discover millions of devices connected to a network.

Recent works have been exploring machine learning models for IoT device fingerprinting.  A common approach is to employ supervised machine learning, to train a single classifier for each device~\cite{Meidan2017ProfilIoT,miettinen2017iot}; and these set of classifiers are subsequently used for predicting a traffic session (or sequence of first $n$ packets) of a device. If a device traffic is predicted to be of multiple types (or classes), then another metric (such as edit distance~\cite{edit-distance-1964}) is used to break the tie. 
In our previous work~\cite{DEFT-2019}, we combined both supervised and unsupervised approaches for identifying known devices as well as grouping unknown devices of the same type across different networks. We explored large number of features extracted from network traffic, and the empirical study based on 15 IoT devices confirmed with other works that, devices can be classified based on network traffic features with high accuracy.

There are also works that consider the cost of including different features, although not in the IoT settings (e.g.,~\cite{ma2008penalized,tavenard2016cost}). We briefly discuss a few here. In~\cite{zhang2015multi} the authors used a particle swarm optimization approach for cost-based feature selection in order to get a Pareto front of non-dominated solutions which gives both high accuracy and low cost. Among other works, Min~{\em et. al} formulated  cost-constrained feature selection as a CSP (constraint satisfaction problem)~\cite{min2014feature}; while they proposed a heuristic solution, they assumed decisions as given, and the notion of risk was not considered within the scope of the work. There are also works that included a cost evaluation function as part of existing feature selection approaches; for example, two filter-based feature selection methods were experimented in~\cite{bolon2014framework}.
%~\cite{chuang2011chaotic}  used particle swarm optimization exploiting its simple nature and fast convergence property. 
%Furthermore, in~\cite{bolon2014framework}, the authors proposed a general solution to the cost-based feature selection problem, by trying to simultaneously optimize the correlation of features with the class and their associated cost. 
Another interesting approach was proposed in~\cite{maldonado2017cost}, wherein the cost of feature selection was made part of the SVM classification model for a credit scoring application.

%In~\cite{tavenard2016cost}, the authors proposed cost-aware early classification of time series by balancing the trade-off between prediction accuracy and detection time.In \cite{ma2008penalized} the authors presented a review of the methodologies which are common in Bioinformatics for penalized feature selection and classification.

However, to the best of our knowledge, previous works did not consider the cost of different features that are extracted for training and classifying devices. Neither did they consider the {\em risk} of misclassification, in particular when the loss due to wrong classification can be different depending on the classification result. We consider these important aspects in our work here.

% In ~\cite{bezawada2018behavioral}, the authors generalized this by periodically cross-verifying the established device fingerprints and hence increased the classification accuracy. Similarly, there has been research work where the devices are classified with respect to the MAC layer traffic of the devices~\cite{siby2017iotscanner}. However, these devices can identify the devices at very high accuracy, none of them take into consideration the cost of extraction of these features, especially the cost of privacy. Also, a feature budget has not been taken into consideration which constrains the number of features the classifier can use for device fingerprinting, depending on several factors. Moreover, the performances across different classifiers has also not been demonstrated though they also play a very important role in the performance of the system.

%\par The cross-entropy method has been widely used for 

%\par Work has also been done on the application of cross-entropy  (CE) technique for optimal selection of features in different fields of application. For example, in \cite{zhang2018spatial}, the authors used the CE for sensor selection. In ~\cite{NADA-2018} the authors developed an algorithm for anomaly detection and attribution in networks. 

%% file: tex/system.tex
\section{System model}
\label{sec:system}

%The system model section should always have a figure.The figure should demonstrate the parameters of your system model. Prepare the figure so that it can later be reused or enhanced to demonstrate your solution.

We describe the system model in this section. Below, we first state the assumptions of the system, and then go on to  define misclassification loss as well as our concept of risk.
The commonly referred notations are listed in Table~\ref{tab:def}. 

\begin{table}[]
\centering
\caption{Table of Notations}
\label{tab:def}
\begin{tabular}{|l|l|}
\hline
Term & Definition \\ \hline \hline
$\mathbf{f}$ & \begin{tabular}[c]{@{}l@{}}Feature vector, constituting of the features abstracting \\ network traffic characteristics. $|\mathbf{f}| = m$\end{tabular} \\ \hline
$\mathbf{c}$ & \begin{tabular}[c]{@{}l@{}}Cost vector, constituting of the costs associated\\  with each of the $m$ features. $|\mathbf{c}| = m$\end{tabular} \\ \hline
$\mathbf{v}$ & \begin{tabular}[c]{@{}l@{}}A binary vector denoting the features selected \\ from $\mathbf{f}$. $|\mathbf{v}| = m$\end{tabular} \\ \hline
$\lambda$ & Feature budget \\ \hline
$R(\mathbf{v})$ & \begin{tabular}[c]{@{}l@{}}Risk score, based on the selected\\ feature vector $\mathbf{v}$\end{tabular} \\ \hline
$\mathcal{L}$ & \begin{tabular}[c]{@{}l@{}}Loss Matrix where element $l_{i,j}$ represents\\ misclassification loss. $|\mathcal{L}| = n \times n$\end{tabular} \\ \hline
\end{tabular}
\end{table}

\begin{comment}

% \usepackage{graphicx}
\begin{table}[]
\centering
\caption{Table of Notations }
\label{tab:def}
\resizebox{\linewidth}{!}{%
\begin{tabular}{|c|l|}
\hline
 Term & Definition \\ \hline \hline
  %$\mathcal{D}$ & Set of Devices, $|\mathcal{D}| = n$ \\ \hline
  %$ q_d$ & Units of IoT network traffic for device $d$\\ \hline
 $\mathbf{f}$ & Feature vector, constituting of the features abstracting network traffic characteristics. $|\mathbf{f}| = m$ \\ \hline
 $\mathbf{c}$ & Cost vector, constituting of the costs associated with each of the $m$ features. $|\mathbf{c}| = m$ \\ \hline
 $\mathbf{v}$ & A binary vector denoting the features selected from $\mathbf{f}$. $|\mathbf{v}| = m$ \\ \hline
$\lambda$ & Feature budget \\ \hline
% $\mathcal M$ & Classification model for training and testing of data \\ \hline
 % $\mathcal X$ & Confusion matrix \\ \hline
% $\mathcal P$ & Performance matrix \\ \hline
 $R(\mathbf{v})$ & Risk score, based on the selected feature vector $\mathbf{v}$ \\ \hline
 $\mathcal{L}$ & Loss Matrix where element $l_{i,j}$ represents misclassification loss. $|\mathcal{L}| = n \times n$\\ \hline
 %$\eta$ & Sample Size of the cross entropy algorithm \\ \hline
 %$\Phi$ & Binary subspace comprising of all possible combinations of $\{0,1 \}^m$ \\ \hline
 %$\gamma$ & Optimal value for $R(\mathbf{v})$ \\ \hline
\end{tabular}%
}
\end{table}

\end{comment}

\subsection{Traffic and features}
 Let $\mathcal D$ denote the set of devices considered for classification in a network; furthermore, let $n = |\mathcal D|$. Corresponding to a device $d \in \mathcal D $, assume $q_d$ units of traffic have been stored and made available for the purpose of device classification. For simplicity, and without loss of generality, we assume $q$ units of traffic are available for every device. Let $\mathcal T^d$ denote the traffic of device $d$. Each unit of traffic (a connection or a session) of $\mathcal T^d$ is processed to extract the {\em feature vector}. We use $\mathbf f$ to represent the $m$ features of interest in our problem: 
  $$ \mathbf f = [f_1, f_2, \dots, f_m ]; \quad \mathbf f \in \mathbb{R}^{m}.$$
 
Denote by $\mathcal E(.)$ the function for feature extraction. For an input traffic data $\mathcal T^d$ of device $d$, $\mathcal E(\mathcal T^d)$ produces a matrix $\mathcal S^d$ of extracted features corresponding to all units of traffic:

\begin{equation}
    \mathcal{S}^d : = \left[\begin{array}{cccc}{s^d_{1,1}} & {s^d_{1,2}} & {\cdots} & {s^d_{1,m}} \\ {s^d_{2,1}} & {s^d_{2,2}} & {\cdots} & {s^d_{2,m}} \\ {\vdots} & {\vdots} & {\vdots} & {\vdots} \\ {s^d_{q,1}} & {s^d_{q,2}} & {\cdots} & {s^d_{q,m}}\end{array} \right],
\end{equation}
\noindent where $s^d_{j, k}$ denotes the $k^\text{th}$ feature extracted from the $j^\text{th}$ traffic unit of device $d$. 
To generalize, we use $\mathbf s$ to denote an extracted feature vector of a device. Obviously, $\mathbf s$ is of length $m$. 
Note that, $\mathcal S^d, \forall d \in \mathcal D$ form the dataset $\mathbb{X}$ used for training and testing the classification model. We use $\mathbb{X}^\texttt{train}$ and  $\mathbb{X}^\texttt{test}$ to denote the partition of the dataset for, training and testing, respectively. 

\subsection{Cost of feature extraction}\label{sec:cost}

Features  characterise different aspects of the network traffic, and are obtained by processing network traffic. Therefore, feature extraction involves cost in terms of resources required for processing and storing the feature, and sometimes for even purchasing the feature. 
With a slight abuse of notation, we use $\mathcal E_{f_k}(.)$ the function to extract feature $f_k, 1 \le k \le m$. Therefore the cost of extracting a feature $f_k$ will be denoted by:
$$ c_k = \texttt{cost}(\mathcal E_{f_k}); \quad \forall k \in [1, 2, \dots, m],$$
 \noindent and the cost vector is defined as:
$$\mathbf c = [c_1, c_2, \dots, c_m]; \quad \mathbf{c} \in  \mathbb{R}_{+}^{m}.$$
  
\subsection{Supervised classification of devices}

 We consider supervised machine learning approaches for IoT classification, wherein traffic data is labelled and provided for training the classifier. The trained model, denoted by $\mathcal M$, is subsequently used in an operational environment to differentiate devices into different types.

Given the matrices of extracted features $\mathcal S^d, \forall d \in \mathcal D$, the classification model $\mathcal M$ maps $\mathbf s$ to one of the devices types $\{d_1, d_2, \dots, d_n\}$. The output of a machine learning model $\mathcal M$ is usually represented using an $n \times n$ confusion matrix $\mathcal X_{\mathcal M}$ that is processed to obtain relevant performance metrics, such as precision and recall. 
%for the specifically trained model $\mathcal M$. 
%For simplicity, we drop the subscript for the model. 
\begin{comment}
\begin{equation}
    \mathcal{X} : = \left[\begin{array}{cccc}{x_{1,1}} & {x_{1,2}} & {\cdots} & {x_{1,n}} \\ {x_{2,1}} & {x_{2,2}} & {\cdots} & {x_{2,n}} \\ {\vdots} & {\vdots} & {\vdots} & {\vdots} \\ {x_{n,1}} & {x_{n,2}} & {\cdots} & {x_{n,n}}\end{array} \right],
\end{equation}
\noindent such that $x_{i,j}$ represents the number of
\end{comment}
The element in the $i^\text{th}$ row and $j^\text{th}$ column of the confusion matrix, $x_{i,j}$, represents the number of data points (or instances) of class $j$ that were predicted to be of class $i$ by the classifier. Given the confusion matrix, we now define $p_{i,j}$, the probability of misclassification, as the probability that an instance of class $j$ is predicted as an instance of class $i$. That is if $\widehat{d}$ is the predicted device and $d$ is the actual device, then we define the probability of classifying $d = j$ as $\widehat{d} = i$ is given as:

\begin{align}
\label{eq:performance}
    {\widehat{p}_{i,j}} = \operatorname{Pr}(\widehat{d}=i | d=j) &  %= \frac{\operatorname{Pr}(\hat{d}=i \cap d=j)}{ \operatorname{Pr}(d=j)} \\ 
    =
     \frac{x_{i,j}}{\sum_{k=1}^{n} x_{k,j}}. & %\forall i,j \in \left[0,1,2, \ldots, J \right] 
\end{align}

For a classification model $\mathcal M$, we define an $n \times n$ misclassification matrix $\hat{\mathcal{P}}_\mathcal{M}$ with elements $\widehat{p}_{i,j}$'s. For simplicity, henceforth, we drop the subscript and denote the misclassification matrix as $\mathcal P$.

\begin{comment}
 \begin{equation}
\hat{\mathcal{P}} :=\left[\begin{array}{cccc}{\widehat{p}_{1,1}} & {\widehat{p}_{1,2}} & {\cdots} & {\widehat{p}_{1,n}} \\ {\widehat{p}_{2,1}} & {\widehat{p}_{2,2}} & {\cdots} & {\widehat{p}_{2,n}} \\ {\vdots} & {\vdots} & {\vdots} & {\vdots} \\ {\widehat{p}_{n,1}} & {\widehat{p}_{n,2}} & {\cdots} & {\widehat{p}_{n,n}}\end{array}\right].
\end{equation}
\end{comment}

\subsection{Misclassification loss} 
\label{subsec:loss}

%As discussed above, confusion matrix captures the performance of a machine learning classifier based on the prediction accuracy. However, the loss due to misclassification is often evaluated in binary---if classification is correct, the loss is zero; otherwise, the loss is one. However, such a binary notion of capturing the classification performances overlooks an important fact \todo{should 've mentioned in Intro; ensure not redundant}. For example, misclassifying a smart bulb of vendor x as a smart bulb of vendor y, might not be as bad as misclassifying the bulb as a camera. In our work here, we would like to capture the losses of misclassification as perceived by users. To this end, we define a misclassification loss matrix, that represents the penalty we would like to assign for different kinds of misclassificatoins due to a particular classifier. 

Assume that the classifier is tasked to predict the class of a device $d_i$. We represent a device $d_{i}$ as as an ordered pair~$<$~\texttt{type}, \texttt{brand}~$>$, where \texttt{type}  indicates the type of the device, e.g., a camera or a speaker, while \texttt{brand} represents the brand of the item, e.g., Sony or Samsung. When the classification model (mis)classifies a device $d_i = <t_i, b_i>$ as $d_j = <t_j, b_j>$, there is a loss incurred; we denote this misclassification loss as $l_{i,j}$, 
%Let us denote the predicted class as $\widehat{d}$, and the actual class as $d$. Then, the loss associated with a decision $\widehat{d}=j | d=i$ is denoted by $l_{i,j} \geq 0$ 
and the corresponding loss matrix $\mathcal L$:

\begin{equation}
\mathcal{L} :=\left[\begin{array}{cccc}{0} & {l_{1,2}} & {\cdots} & {l_{1,n}} \\ {l_{2,1}} & {0} & {\cdots} & {l_{2,n}} \\ {\vdots} & {\vdots} & {\vdots} & {\vdots} \\ {l_{n,1}} & {l_{n,2}} & {\cdots} & {0}\end{array}\right].
\end{equation}

This loss matrix is a user-defined parameter and depends on the types of devices considered for classification. While this allows for a generic definition of the misclassification losses as deemed right by the user, we later provide (in Section~\ref{subsec:loss-eval}) the specific definition used in this work for the purpose of illustration as well as for experimentation.

\subsection{Cyber Risk}
\label{subsec:risk}
%\todo{Biswadeep: define $l_{i,j}$}

Using these definitions, we define the cyber risk of the model. %which is the main objective function of our paper.
While $\mathbf f$ is the vector of features of interest, not all features might be available for modeling, given features have cost, and a solution is constrained by the available budget. We use a binary vector $\mathbf{v}$ to indicate whether a feature is selected or not. 
\[\mathbf{v}=\left[v_{1}, \dots, v_{m}\right],\] 
where
\[v_{k}=\left\{\begin{array}{ll}{1,} & {\text {if the } k^{\text {th }} \text { feature selected }} \\ {0,} & {\text {if the } k^{\text {th }} \text { feature is not selected }}\end{array}\right.\]

%Our objective is to find the optimal $\mathbf{v}$ which minimizes the Cyber Risk, $R(\mathbf{v})$ At the same time we also aim to keep the cost of the selected features, given by the cost vector $\mathbf{c}$ within a specified feature budget, $\lambda$.\\
Next, we define cyber risk. Given a classifier $\mathcal M$, loss matrix $\mathcal L$, feature budget $\lambda$, the extracted data (for the  selected features indicated by $\mathbf v$) $\mathbb X_\mathbf{v}$, and a feature selection vector $\mathbf v$, the cyber risk score $R(\mathbf{v},\mathcal{M}, \mathcal{L}, \lambda, \mathbb X_\mathbf{v})$ is the expected sum of the losses due to misclassifications. Since we are considering the risk score for a fixed classifier model, for a fixed number of devices and traffic data, we may simplify the risk score by denoting it as $R(\mathbf{v})$.
%If we assume the predicted class from the classifier as $j$ and the actual class to be $i$, then the probability of misclassification is given by $\operatorname{Pr}(\widehat{d}=i | d=j)$. \\
Formally, the \textbf{cyber risk score} or the \textbf{risk score}, based on the set of selected features  (captured by $\mathbf{v}$), is defined as the sum of the product of the probability of misclassification  %$\operatorname{Pr}(\widehat{d}=i | d=j; \mathbf{v})$ 
and the losses associated with misclassification of each device ($l_{i,j}$), which is given by:
\begin{align}
    R(\mathbf{v}) &= \sum_{i=1}^{n} \sum_{j=1}^{n} \operatorname{Pr}(\widehat{d}=i | d=j; \mathbf{v}) \times l_{i,j} 
    \label{eq:cyber_risk}
\end{align}

\section{Problem Definition - Optimal Feature Selection under Budget Constraint} \label{sec:problem}
%Often, this section is merged with the system model. State your problem clearly. Be as exact as possible into stating what the question of the problem is. It reflects poorly upon an author if he cannot describe or does not know what problem his solution addresses. But most importantly, it will be easier for successive researchers to classify your work

This work looks into the problem of minimizing the cyber risk associated with classification of IoT devices, by selecting the optimal set of features from a universal feature set, whilst keeping the feature cost less than the budget available.  \\

%\begin{probdef}
\noindent {\bf Problem Definition:} {\em Given the misclassification matrix $\mathcal{P}$ for a given feature vector $\mathbf{f}$, and the loss matrix $\mathcal{L}$, we aim to find the optimal feature vector (via $\mathbf{v}$) which minimizes the cyber risk score $R(\mathbf{v})$, under a budget constraint $\lambda$.} \\
%\end{probdef}

\noindent Thus, the optimization problem may be defined as:
\begin{comment}
\begin{equation}
    \begin{aligned} \widehat{\mathbf{v}}=& \underset{\mathbf{v} \in\{0,1\}^{m}}{\arg \min } R(\mathbf{v}) \\  \text { s.t } & \mathbf{c}^{T}\mathbf{v} \le \lambda \end{aligned}
\end{equation}
\end{comment}

\begin{equation}
\label{eq:optim}
    \begin{aligned} \mathbf{v}_{s}=& \underset{\mathbf{v} \in\{0,1\}^{m}}{\arg \min } R(\mathbf{v}) \\  \text { s.t } & \sum_{j=1}^{m} {c}_{j} v_{j}\le\lambda,
    \end{aligned}
\end{equation}

%1.Explain R(\mathbf{v})
%2. This is a hard problem because R(\mathbf{v}) us not analytic and v is not convex.
%3.Brute force method solution

\noindent where $R(\mathbf{v})$ is as defined in Eq.~\ref{eq:cyber_risk}. 

%\par The merit of the above problem lies in the fact that the risk function $R(\mathbf{v})$ is not an analytic function i.e., we do not have a definite function representing the behavior of the function. Moreover, the constraint equation used is also not convex. Hence, a combinatorial solution is the only possible solution to the problem. General method of solving such a combinatorial problem, using the brute force method,discussed in Section \ref{VB}, is computationally very expensive. So, in this paper we aim to solve this problem by proposing a novel method of solving the said combinatorial optimization problem. 

The optimization problem presented in Eq. \ref{eq:optim} has some unique properties making it a non-trivial problem to solve. Firstly, the cyber risk function $R(\mathbf{v})$ used as the principle objective function, is not an analytic function. Hence, using methods to solve continuous optimization problems, like the gradient descent is not feasible in this scenario. This is mainly because the gradient descent uses the Newton's Method to find the optimal solution which in turn uses a well-defined continuous objective function; and $R(\mathbf{v})$ is not a continuous function. 
Moreover, another important constraint to use such methods requires the objective function, the constraint equations and the domain space to be convex in nature. In our scenario, we do not know whether we have a closed-form analytical expression for the risk score which is the objective function. This is because it is dependent on the classification model and the features selected. Though a probabilistic model for a simple classifier like the Naive-Bayes classifier can be obtained, we cannot say the same for any general classifier, for example the Decision tree or the Random Forest classifiers. Thus, in a general sense, it is not possible to get an analytical expression for the risk score defined in Eq. \ref{eq:cyber_risk}. The best we can do is to obtain a point-wise evaluation of the objective function which is the minimum requirement to attempt a solution to this optimization problem.
We observe that the constraint functions are affine in nature, and since all affine functions are convex, the constraint functions are also convex in nature. This leaves us with the domain space ($m$-dimensional binary subspace) for the optimization problem.
%which is $m$-dimensional binary subspace $\Phi$.%\todo{this sentence is not clear; please rephrase}.
\begin{comment}

\begin{lemma}
A set of $K$-dimensional Binary subspace is not convex in nature
\end{lemma}
\textbf{Proof:} Let us consider a $m$-dimensional binary subspace represented by $\Phi \in \{0,1 \}^m$ is convex in nature. That implies that the line segment between any two points in the set $\Phi$ , i.e., $\lambda_1, \lambda_2 \in \Phi$  and any $\theta$ with $0 \le \theta \le 1$ we will have
\[ \theta \lambda_1 + (1-\theta) \lambda_2 \in \Phi \]

Let us consider the case when we select  the maximum number from the $m$-dimensional binary subspace as one of the points to be considered: 
$ \overbrace{1111\dots111}^{m- \text{digits}}$
%We select the other point as any other number other than lowest number given by all zeroes, and then take the linear combination of the two with $\theta \neq 0 or 1$.
The second point is chosen as any point other than the vector constituting of all zeroes. We then take the linear combination of the two points with $\theta \neq 0 \hspace{0.1 cm} \text{or} 1$
Then the resultant point will be in the $(m+1)$-dimension binary subspace which does not belong to $\Phi$, the $m$-dimensional binary space of our problem.\\
Hence, we can conclude that a $m$-dimensional binary subspace $\Phi$ is not a convex set.\\
\end{comment}
\par Since the domain space is not  convex in nature, we can safely conclude that the problem is not a convex optimization problem, and cannot be solved by using convex optimization tools. Therefore, the optimal way to solve such problems is via a combinatorial search over all possible combinations of features by the brute force method which is not a very efficient way.  In classes of problem such as the one posed in this paper,a subsequent brute force enumeration of all possibilities is computationally prohibitive and as such a stochastic search based approach needs to be developed in practice. In the next section, we explore the cross entropy technique for searching the entire subspace in an efficient way to potentially converge to the global optima.

%% file: tex/algos.tex
\section{Feature Selection via Cross Entropy Method}
\label{sec:cross}
%We need to find the optimal set of features within our budget, so that we can classify the IoT devices with the least misclassification loss. In order to do this, we used the cross entropy based optimization method to find the solution set. When compared to other method like the brute force, the cross entropy is converges at a much faster pace and is also computationally much less expensive, as shown in Section IV. The main motivation behind choosing cross entropy method is because of its simple nature.  

In this section we develop a novel algorithm to solve the feature selection under budget constraints, presented in Eq.~\ref{eq:optim}. As mentioned before, this is a non-convex combinatorial optimization problem. To overcome this computational difficulty, we develop a novel algorithm via a Monte Carlo sampling approach, which has a low computational complexity. Specifically, we use the cross entropy (CE) technique, which is based on the Importance Sampling Technique which is known to be an effective solution in computational approaches to otherwise solving NP-hard problems.
The CE converts the deterministic optimisation problem into a stochastic couterpart, see details in~\cite{asmussen2005heavy}. 
It then approximates the optimal sampling distribution by minimizing the Kullback-Leibler (KL) divergence~\cite{de2005tutorial}. 
Without loss of generality, we may change the optimization problem in Eq. \ref{eq:optim} to a maximization problem, that is: 
\begin{equation}
\mathbf{v}^* = \underset{\mathbf{v} \in\{0,1\}^{m}}{\arg\max} \quad \mathcal{U}(\mathbf{v}), \quad \text{s.t.}\quad \mathbf{c}^T\mathbf{v} \le \lambda
\end{equation}
where $\mathcal{U}(\mathbf{v}) = \frac{1}{R(\mathbf{v})}$.
%The CE technique solves the optimization problem by estimating %rare-event possibilities and hence, locate an optimal parametric %sampling distribution.

Instead of searching for the optimal solution directly, like in the case of a brute force search based approach, the CE places a probability distribution on the set of  $m$-dimensional binary  features $(\mathbf{v} \in \{0,1 \}^m)$, thus transforming the problem into estimation.

%To do this, we define a set of indicator functions given as %$\mathbbm{1}(\mathcal{U}(\mathbf{v}) \ge \gamma)$ on $\Phi \in \{0,1 \}^m$ %for various levels of threshold values $\gamma \in \mathbb{R}$. Next, let %$\{f(\cdot ; \mathbf{\theta}), \mathbf{\theta} \in \Theta \}$ be a family of %probability mass functions on $\Phi \in \{0,1 \}^m$ which is parameterized %by a real-valued parameter vector $\theta$. For a fixed $u \in \Theta$, we %associate the problem of rare-event probability with the optimization %problem as follows:
%\begin{equation}
%    \mathbb{P}_u(\mathcal{U}(\mathbf{v}) \ge \gamma) = %\mathbb{E}_u\left[\mathbbm{1}(\mathcal{U}(\mathbf{v}) \ge \gamma) \right]
%\end{equation}
%where $\mathbb{P}_u$ is the probability measure under which the random state %$\mathbf{v}$ has a discrete probability mass function  $f(\cdot ; \theta)$, %$\mathbb{E}_u$ denotes the corresponding expectation operator. 
In our model, since the selection of a feature is indicated by a binary variable (``select''  or  ``don't select''), 
%($0 \rightarrow$ not selected; $1 \rightarrow$ selected), 
we use an independent Bernoulli random variable to indicate this choice for each of the $m$ features. The Bernoulli distribution
has a single variable parameter~$p$ , i.e., for the $k$-th feature this is given by $\{p_k\}_{k=1}^m$, and is a member of the Natural Exponential Families (NEF) of distributions \cite{collins2002generalization}. This is a big advantage, since under the NEF, the parameter of the distribution can be estimated analytically in closed form via the Maximum Likelihood Estimator (MLE), making the CE easy to implement. 
At each iteration of the CE method there are three steps:
\begin{enumerate}
    \item[S1] Generate samples: generate $\eta$ independent samples of binary sets given by $\mathbf{w}^{[i]} = \{w^{[i]}_{1},w^{[i]}_{2} \ldots, w^{[i]}_{m}\}$, where $w^{[i]}_{k} \sim \texttt{Bernoulli}(p_{k}) ; 1\le i \le \eta$.

    \item[S2] Selection of elite samples: from those $\eta$ samples, we select only those samples which satisfy the budget constraint  (ie. $\mathbf{c}^T\mathbf{w}^{[i]} \le \lambda$) to obtain a subset of $\eta' \leq \eta$ samples. 
    Next, from this subset we select only the ``elite'' samples---those for which the objective function value exceeds a pre-defined threshold, defined via a quantile value $\rho$. This choice is indicated by 
    $\mathbbm{1}\left(\mathcal{U}(\mathbf{w}^{[i]}) \geq \gamma\right)$,
    where $\gamma$ is the $(1- \rho)^{\text{th}}$-sample quantile of $\mathcal{U}_{1:\eta'}$.
    \item[S3] Estimation of parameters: using only this subset of elite samples, we re-estimate the parameters of the Bernoulli random variable $\{p_k\}_{k=1}^m$ via the MLE, given by
\begin{multline}
\widehat{p}_{k}=  \frac{\sum\limits_{i=1}^{\eta'} \mathbbm{1}\left(\mathbf{w}^{[i]}_{j} =1\right) 
\overbrace{
\mathbbm{1}\left(\mathcal{U}(\mathbf{w}^{[i]}) \geq \gamma\right)}
^{\text{Choose elite samples}}
}{\sum\limits_{i=1}^{\eta'} \mathbbm{1}\left(\mathcal{U}(\mathbf{w}^{[i]}) \geq \gamma\right)}, \\
k=\{1,\cdots,m \} .
\end{multline}
\end{enumerate}
\noindent 
The three steps of the algorithm are then iterated until a stopping rule is met, for example, if the algorithm has converged to a local maxima or has exhausted a pre-defined number of iterations.
The parameter $\rho$ is the quantile value, and is commonly set to $0.9$. 
%We also choose a learning rate $\alpha \in [0,1]$ to update the value of $v_{j}$ from its previous iteration using a weighted average formulation. 
 The resulting algorithm is presented in Algorithm 1, where in Step 10 we have introduced the parameter $\alpha \in [0,1]$ which controls the learning rate of the algorithm, thus avoiding getting trapped in local maxima, see details in  \cite{de2005tutorial}.

It is important to note that $v_{k}$,
where $1 \le k \le m $, denotes whether the $k^\text{th}$ feature is selected by the algorithm. 
%On the other hand, $w_{i}^{[j]}$ represents the  $i$-th value of the vector $\mathbf{w}$ for the $j$-th sample. Thus, $1 \le i \le m , 1 \le j \le \eta$, where $\eta$ is the sample size. 
We iterate the cross entropy algorithm until the convergence criterion is met, which in this work, translates to reaching the maximum number of iterations ($T_{\max}$). 
 %to obtain the optimal feature selection vector $\mathbf{v}_{s}$.

\begin{algorithm} 
\label{alg:cross}
\SetAlgoLined
  \KwInput{ $\mathbb{X}^\texttt{train},  \mathbb{X}^\texttt{test}, \eta, \rho \in [0,1), \alpha, \beta, \lambda, \mathbf{c}$} 
  
  %\KwData{Testing set $x$}
  
   %$t \leftarrow 0$  \Comment{initialize iteration number}
  
  %t \leftarrow 0 \Comment{Initialize Iterator}
  
  Initialize vague prior: $\mathbf{\widehat{p}} = \left[\widehat{p}_{1}, \widehat{p}_{2}, \ldots , \widehat{p}_{m}\right]$ such that $p_{k} = 0.5 , \forall k$ \\
  \While{\texttt{stopping criterion not met}} 
    {
    \underline{Step 1:}\\
        Generate $\eta$ independent samples of binary sets given by 
        $\mathbf{w}^{[i]} = \left[w^{[i]}_{1}, w^{[i]}_{2}, \ldots, w^{[i]}_{m}\right]$, where $w^{[i]}_{k} \sim \texttt{Bernoulli}(\widehat{p}_{k}) ; \quad 1\le i \le \eta, 1\le k \le m$.
         
        %For each of the $\eta$ samples, compute  $\mathcal{U}_{i} = \mathcal{U}(\mathbf{w}^{[i]}), i = {1 , \ldots, \eta}$ as given in Eq: (\ref{eq:risk}),
        %where $v_{j} =\left\{\begin{array}{ll}{1,} & {w^{j}_{1} = 1} \\ {0,} & {\text { Otherwise }}\end{array}\right., \forall v_{j} \in \mathbf{v}^{m}$.\\
        \underline{Step 2:}\\
         From $\eta$ samples, remove all samples that do not satisfy the constraint $\mathbf{c}^{T}\mathbf{w}^{[i]} < \lambda$. Let the new indices be $i = {1 , \ldots, \eta'}$.
            
         %From $\eta$ samples, store all $i$ that satisfies the constraint $\mathbf{c}^{T}\mathbf{w}^{[i]} < \lambda,\quad i = {1 , \ldots, \eta'}$

    $\mathcal{X} = \mathcal{M}(\mathbb{X}^\texttt{train}_{\mathbf{w}^{[i]}} , \mathbb{X}^\texttt{test}_{\mathbf{w}^{[i]}})$ 
     
     Compute $\hat{\mathcal{P}}$, from $\mathcal{X}$ as given in Eq. \ref{eq:performance}. 
    
    Compute  $\mathcal{U}_{i} = \mathcal{U}(\mathbf{w}^{[i]}), i = {1 , \ldots, \eta'}$.

        Compute $\gamma$, the $(1- \rho)^{\text{th}}$-sample quantile of $\mathcal{U}_{1:\eta'}$. 

        %$v_{k} \leftarrow \alpha \frac{\sum\limits_{i=1}^{\eta'} %\mathbbm{1}\left(\mathbf{w}^{[i]}_{k} =1\right) %\mathbbm{1}\left(\mathcal{U}(\mathbf{w}) \geq %\gamma\right)}{\sum\limits_{i=1}^{\eta'} %\mathbbm{1}\left(\mathcal{U}(\mathbf{w}) \geq \gamma\right)}$ + %$(1-\alpha)v_{k}$, $\forall  k$

\underline{Step 3:}\\
     Re-estimate model parameters via the MLE:  
       $
\widehat{p}_{k}= 
\alpha
\frac{\sum\limits_{i=1}^{\eta'} \mathbbm{1}\left(\mathbf{w}^{[i]}_{j} =1\right) \mathbbm{1}\left(\mathcal{U}(\mathbf{w}^{[i]}) \geq \gamma\right)}{\sum\limits_{i=1}^{\eta'} \mathbbm{1}\left(\mathcal{U}(\mathbf{w}^{[i]}) \geq \gamma\right)} +(1-\alpha)\widehat{p}_{k}$, $\forall  k$

        %$t = t+1$
    }

    For each $p_k$, make the final binary decision as follows:
        $v_{k} =
                \left\{\begin{array}{ll}{1,} & {p_{k} \geq \beta} \\ {0,} & {\text { Otherwise }}\end{array}\right. \forall k \in \left[1,\dots, m \right]$
        where $\beta$ is a pre-defined threshold.
        \\

    \Return $\mathbf{v}^* = \left[v_{1}, v_{2}, \ldots , v_{m}\right]. $

\caption{Cross Entropy Based Feature Selection}
\end{algorithm}

\par

\textbf{Computational Complexity:}
The computational complexity of the CE algorithm for feature selection is characterized by the number of independent samples used $\eta$, the feature size $m$ and the stopping criterion. In Algorithm~1, the calculation of $\mathcal{U}(\mathbf{w}^{[i]})$ is the most computationally expensive step. This is because the computation of $\mathcal{U}(\mathbf{w}^{[i]})$ involves using a classifier and matrix multiplication of the misclassification matrix and the loss matrix. Thus, this step becomes the bottleneck for the performance of the  algorithm, which leads us to conclude that the complexity of computation of this step will also be the dominant component of the complexity of the entire algorithm. Now, the complexity of computation of the $\mathcal{U}(\mathbf{w}^{[i]})$ is given by $\mathcal{O}(\eta \times \mathcal{C}(m; \Psi) \times T_{\max})$ where $T_{\max}$ is the number of iterations and $\mathcal{C}(m; \Psi))$ denotes the complexity of the classifier in terms of the feature length $m$ and other parameters $\Psi$ specific to the classifier. For example, if we use the Naive-Bayes classifier, $\mathcal{C}(m; \Psi) = \mathcal{O}(N_{\texttt{train}}m)$, where $N_{\texttt{train}}$ is the number of training examples and $m$ is the number of features. Since the number of samples is a constant for each iteration, the dependent parameters for the complexity of the algorithm may be reduced down to $\eta$ and $T_{\max}$, which are also the only free parameter that can be controlled by us. The values of $\eta$ and $T_{\max}$ are determined based on the trade-off between the computational budget and the required detection performance.

%\par For example, if we use the Naive-bayes classifier, $\mathcal{C}(m; \Psi) = \mathcal{O}(N_{\texttt{train}}m)$, where $N_{\texttt{train}}$ is the number of training examples and $m$ is the number of features. Thus, the overall complexity of the cross entropy algorithm is formulated as $\mathcal{O}(\eta T_{\max} \times N_{\texttt{train}}m)$. On the other hand, if we use the Decision Tree based calssifier, which has a complexity of $\mathcal{O}(N_{\texttt{train}}m\log(N_{\texttt{train}}))$, the overall complexity becomes $\mathcal{O}(\eta T_{\max} \times N_{\texttt{train}}m\log(N_{\texttt{train}}))$. 

\section{Feature selection via  brute-force and greedy approaches} \label{sec:others}

In this section we develop algorithmic alternatives to the CE technique that will act as comparisons to our proposed approach. %These will serve as comparison to CE from a few perspectives, such as performance and computational complexity. 
We first present the brute-force approach for solving the optimization problem (Eq.~\ref{eq:optim}). 
%This method solves the optimisation problem optimally and serves as a lower bound for cyber risk, at the expense of very high computational complexity. 
Subsequently, we also develop three greedy heuristics that use different strategies for selecting features for device classification. 

%In this section, we compare five different approaches to solve the above optimization problem. At first, we use the brute force approach to evaluate the cyber risk score for each and every possible feature combination. This gives us a upper bound on the run-time complexity of the algorithm, while giving the lower bound on the risk. Further on, we implemented three different greedy algorithms for the classification problem, with respect to the feature cost, the risk score and the ratio of accuracy to feature cost.
\par 
%From the complete dataset of $m$ features, we derived the training set and the testing set using a cross validation technique. For simplicity, we represent the training set using the feature vector $\mathbf{v}$ as $\mathbb{X}^{\texttt{train}}_{\mathbf{v}}$ and the testing set as $\mathbb{X}^{\texttt{test}}_{\mathbf{v}}$. We use the classifier model $\mathcal{M}$ for the classification purpose to obtain the confusion matrix $\mathcal{X}$ from which the Performance Matrix $\hat{\mathcal{P}}$ is derived.

\subsection{Brute Force Method}\label{VB}

The brute force approach is presented in Algorithm~\ref{alg:brute}. The algorithm selects each combination of the features (via the vector $\mathbf v$), and evaluates the risk score if the cost of the features is within the budget constraint ($\lambda$). Of all such feature combinations, one that minimizes the risk score is the optimal set of features for device classification.
Though this method gives the optimal solution over the feature space, the computational complexity of searching over the entire parameter space exhaustively is prohibitively high. Thus, this approach gives us an upper bound on the run-time complexity of the algorithm, and also a lower bound on the risk.

\par \textbf{Computational Complexity:} The exhaustive search in the algorithm means that, for a feature vector of length $m$, the \texttt{while} loop is executed $2^{m}$ times. 
For each of the $2^m$ executions, a classifier has to be trained (and tested) to determine the loss associated with the selected feature vector. Thus, the overall complexity of the brute force search algorithm is given by $\mathcal{O}(2^{m} \times \mathcal{C}(m; \Psi))$.

\begin{algorithm}[t]
\SetAlgoLined

  \KwInput{$\mathbb{X}^\texttt{train}, \mathbb{X}^\texttt{test}$, $\mathbf{c}$, $\lambda$, $\mathbf{f}$, $\mathcal{M}$,$\mathcal{L}$}
  
  $i \leftarrow 1 , R_{\min} \leftarrow \infty$ \Comment{initializations}\\
  $\mathbf{v} \leftarrow \left[0,  \dots, 0\right]_{(1 \times m)}$ \\
    \While{ $i \le 2^{m}$} 
    {
        $\mathbf{v} \leftarrow \texttt{Binary}(i)$  \Comment{$v_{k}$ is $k$-th digit in the Binary representation of $i, \quad 1 \le k \le m$ }
            
        \If{$ \mathbf{c}^{T}\mathbf{v} < \lambda$}
            {

    $\mathcal{X} \leftarrow \mathcal{M}(\mathbb{X}^\texttt{train}_{\mathbf{v}} , \mathbb{X}^\texttt{test}_{\mathbf{v}})$ 
     
     Compute $\hat{\mathcal{P}}$, from $\mathcal{X}$ as given in Eq. \ref{eq:performance}

    Compute $R(\mathbf{v})$ according to Eq. \ref{eq:cyber_risk} 
            
     \If{$R(\mathbf{v}) < R_{\min}$}
     {
     $R_{\min} \leftarrow R(\mathbf{v})$ \\
     $\mathbf{v}_{s} \leftarrow \mathbf{v}$ \\
     }
             }
            $i \leftarrow i+1$ \\
    
    }
    \Return $\mathbf{v}_{s}$
\caption{Brute Force Feature Selection}
\label{alg:brute}
\end{algorithm}

%\par The brute force algorithm is helpful as it helps us to identify the global optima of the loss function, which is critical for the validation of our proposed method, especially since we do not have an analytical solution to the problem.
%Moreover, the solution also gives an upper bound on the computational complexity, as the proposed algorithm should have a complexity lesser than that of the brute-force algorithm in order for it to have any significance.

%\If{$\sum_{j} c_{j} \times {v}_{j} < \lambda$}

\subsection{Greedy Algorithms}

As the name suggests, greedy algorithms are efficient ways to solve an optimization problem, with the down side being that, they might end up with a local optimum. We define three intuitive greedy approaches in this section; they differ on the parameter (referred to as \textbf{\texttt{key}}) they used to select the next feature for classification. The general steps for these three greedy approaches are given in Algorithm~\ref{alg:greedy}. First, the feature indices are sorted based on an input parameter \textbf{\texttt{key}}. Subsequently, in each iteration within the \texttt{while} loop, the algorithm selects one feature at a time, constrained by the budget, thereby eventually forming the feature vector (indicated by $\mathbf v$) for classification.

%The greedy algorithm always makes a locally optimum choice at each step and tries to reach the global optima by doing so. In this paper, the greedy algorithm selects the features with respect to the parameter of choice at each step like the feature cost, referred to as the "weight", the cyber risk score, referred as "value" or the accuracy to cost ratio, referred as the "density". In the greedy approach, the first step involves obtaining the arguments of the sorted feature list according to a specific parameter. Thereafter, from this sorted array the features are selected one at a time, if the cost of addition of the feature does not surplus the specified feature budget. 
%Eg., if we are selecting the features with respect to feature cost, then the greedy algorithm will select the feature with the least cost, followed by the one with the second-lowest cost and so on until either the feature budget is exceeded or the entire feature set is exhausted, in order to maximize the number of features selected. This is based on the hypothesis that a greater number of selected features will lead to a better overall classification result.  
%The greedy algorithm may be considered of three main types:

\subsubsection{Cost-based greedy algorithm (CGA)}
    %In this approach, we have only considered the cost of the features as the parameter to select the features. The features are first ranked in ascending order of the feature cost after which a greedy algorithm is used to select the features starting from the one with the least cost, and going on until the allotted feature budget is exhausted. This is based on the hypothesis that a greater number of selected features will lead to a better overall classification result.
    
    In this approach, the feature cost ($\mathbf c$) is used as the input parameter \textbf{\texttt{key}} to sort the (indices of) features. Therefore, this simple greedy algorithm selects the minimal cost feature at each iteration. This would reveal the performance of classification, if only feature cost is used as a criterion to select the features. \\ %This process is based on the hypothesis that a greater number of selected features will lead to a better overall classification result.\\
    \textbf{Complexity:} The sorting algorithm takes $\mathcal{O}(m\log m)$ time for $m$ features. The classifier has to be trained only once in this case, after the selection of all the features. Hence, the complexity of this algorithm is $\mathcal{O}(m\log m + \mathcal{C}(m; \Psi))$.

\subsubsection{Risk-based greedy algorithm (RGA)}
   % In this method, we used the greedy algorithm based on the cyber risk score of the classification using the features. In order to do so, firstly, we calculated the risk using only one feature at a time. Thus, we obtained a 'value' associated to each of the features and ranked them in the descending order of priority, with the feature giving the least risk individually being ranked the first. Then, the greedy algorithm is used to select the features of different costs, ranked according to the risk score of each, until the allotted feature budget is exhausted. Thus, in this method we get the set of features with the minimum sum of risk scores within the feature budget. It is based on the principle that if we select the features giving the least risk scores individually, then the combination of those features will give a good approximation to the global minima of cyber risk score. \\
   Next, we define a risk based approach. We train a classifier using only one feature at a time, and we do this for all features constituting~$\mathbf f$. For each classifier corresponding to each feature $f_i, 1 \le i \le m$, we compute the risk using Eq.~\ref{eq:cyber_risk}. The vector of risk scores of length $m$ is then provided as input parameter \textbf{\texttt{key}} to Algorithm~\ref{alg:greedy}. Therefore, RGA algorithm attempts to select the features that correspond to the least risk scores, under the assumption that the risk due to (say) two features is equal to the sum of the risk due to the individual features. \\
   %In this method, we used the greedy algorithm based on the cyber risk score of the classification using the features. In order to do so, firstly, we calculated the risk using only one feature at a time. Thus, we obtained a 'value' associated to each of the features and ranked them in the descending order of priority, with the feature giving the least risk individually being ranked the first. Thus, in this method we get the set of features with the minimum sum of risk scores within the feature budget. It is based on the principle that if we select the features giving the least risk scores individually, then the combination of those features will give a good approximation to the global minima of cyber risk score. \\
    \textbf{Complexity:} The classifier is run $m$ times to build $m$ models, and obtain their corresponding risks. Therefore, the overall complexity is $\mathcal{O}(m\log m + m \times \mathcal{C}(m; \Psi))$

\begin{algorithm}[t]
\SetAlgoLined

    \KwInput{$\mathbb{X}^\texttt{train}, \mathbb{X}^\texttt{test}$, $\mathbf{c}$, $\lambda$, $\mathbf{f}$, $\mathcal{M}$, $\mathcal{L}$, \textbf{\texttt{key}}}
  
  $\Phi \leftarrow \arg \texttt{sort}(\mathbf{f}$, \textbf{\texttt{key}}) \Comment{Obtain the sorted indices}\\
    $i \leftarrow 1$\\
    $\mathbf{v} \leftarrow \left[0, \dots, 0\right]_{(1 \times m)}$ \\
    \While{$i \le m$}
    {
    
    $\mathbf{v}_{\left[\Phi(i) \right]} \leftarrow 1$ \Comment{$\mathbf{v}_{\left[j \right]}$ denotes the $j$-th index of $\mathbf{v}$}
    
    \If{$ \mathbf{c}^{T}\mathbf{v} > \lambda$}
    {
    $\mathbf{v}_{\left[\Phi(i) \right]} \leftarrow 0$
    }
    $i \leftarrow i+1$\\
    }
    $\mathcal{X} \leftarrow \mathcal{M}(\mathbb{X}^\texttt{train}_{\mathbf{v}} , \mathbb{X}^\texttt{test}_{\mathbf{v}})$ 
     
     Compute $\hat{\mathcal{P}}$, from $\mathcal{X}$ as given in Eq. \ref{eq:performance}

    Compute $R(\mathbf{v})$ according to Eq. \ref{eq:cyber_risk} 

    \Return $\mathbf{v}$
\caption{Greedy Feature Selection}
\label{alg:greedy}
\end{algorithm}

\subsubsection{Value-based greedy algorithm (VGA)}
    Finally, we also consider how valuable a feature is based on both the accuracy and the feature cost. 
    %We have also implemented a feature selection algorithm using a Accuracy-Cost ratio-based greedy approach. 
    For each feature $f_i \in \mathbf{f}$, we train the classifier similar to what we did in the RGA approach. The $\text{F}_1$ score is computed for each of $f_i$ and is denoted by $\texttt{acc}(f_i)$. As defined in Sec. \ref{sec:cost}, $c_i$ denotes the cost of feature $f_i$. Thus, we define the value of a  feature $f_i$ as: \[\texttt{density}_i = \frac{\texttt{acc}(f_i)}{c_i}, \quad 1 \le i \le m\]
    The vector of values is then passed as the parameter \textbf{\texttt{key}} to Algorithm~\ref{alg:greedy}.

    %The density was calculated by dividing the classification accuracy by the cost of the feature i.e.,
    %\[\text{Density} = \frac{\text{Classification Accuracy}}{ \text{Cost of the feature}} \]
    %Then, the greedy algorithm is used to select the features from this set based on the density calculated. \\
    %It is based on the principle that we can simultaneously select the features which give the minimum cyber risk score and feature cost.
 %Thus, the features are ranked according to the descending order of the cost, and thus the features which simultaneously gives maximum accuracy to cost ratio is given the maximum priority in selection procedure.\\
  \textbf{Complexity:} Like the previous greedy algorithm, in this algorithm too, the classifier is run $m$ times to train the model for the $m$ features independently. Therefore, the complexity is $\mathcal{O}(m\log m + m \times \mathcal{C}(m; \Psi))$.

%\textbf{Computational Complexity:} The computational complexity of a greedy algorithm, if we are using a feature set of size $f_N$, will be the same as the complexity of sorting the features based on the required key value. This is because for each iteration, the greedy algorithm requires $\mathcal{O}(f_N)$. The sorting process requires $\mathcal{O}(f_Nlog f_N)$,since we are using quicksort algorithm[]. Thus the overall complexity of the greedy algorithm is given by $\mathcal{O}(f_N)$

{
}

%% file: tex/perf-eval.tex
\section{Performance Evaluation}\label{sec:simulations}
%In this section we discuss about the experiments we performed and the datasets and models used for doing so.

In this section, we describe the experiments performed to evaluate the different algorithms presented in the previous sections. After a briefing on the dataset used, we describe how we estimated costs of features (in Section~\ref{subsec:feature-costs}). We also define the loss matrix used for experiments in this work (Section~\ref{subsec:loss-eval}). The first set of experiments we carry out (in Section~\ref{subsec:compare-classifiers}) are to evaluate a selected set of classification models. Subsequently, we evaluate brute force, CE and greedy algorithms, first under no budget constraints (Section~\ref{subsec:eval-no-budget}), and then later we evaluate CE and greedy algorithms when budget is fixed (Section~\ref{subsec:eval-budget}). In Section~\ref{subsec:eval-budget}, we also perform further analysis to compare CE algorithm with the best greedy algorithm. 
%To do so, we collected data from different IoT devices and extracted the features from this raw dataset. We defined a cost of extraction of the features and a loss matrix for classification. 
%\par At first we compared the performance of the different algorithms without any budget constraint. Following this, we implemented the budget constraint and compared the performance of the algorithms discussed. After that, we perform a comprehensive comparative analysis for the performance of our proposed algorithms with respect to the feature size and the feature budget. At last, we compare the variation in performance due to the choice of the classifier. 

\subsection{Real datasets for experiments}
%We have collected the data from 15 different IoT devices 
%write stuff from the other paper about the dataset The dataset was divided into the training set and set using a k-fold cross validation technique with $40\%$ of the dataset considered as the testset.
For the experiments, we used network traffic from 15 different IoT devices as listed in Table I. The devices were connected to the Internet via a gateway, where the network traffic was captured. The traffic was split into intervals of 15 minutes each, and the number of sessions captured for each device is shown in Table~\ref{tab:devices}. From the network data of the IoT devices, 111 features were extracted (see~\cite{DEFT-2019} for the list of features). We make the dataset publicly available~\cite{iot-dataset}.
%, out of which 70 are continuous numeric type features. \todo{Biswadeep: are we taking these 70 numeric types for our experiments? If so, mention here.}
%that could be used for device classification. 

\begin{table}[t]
\centering
\caption{Information on IoT devices used}
\label{tab:devices}
%\resizebox{\textwidth}{!}{%
\begin{tabular}{llll}
\hline
Label & Device & Brand  & \begin{tabular}[c]{@{}l@{}}Sessions\\ Captured\end{tabular} \\ \hline
1 & Echo Dot & Amazon  & 490 \\ \hline
2 & Smart Remote & Broadlink & 480 \\ \hline
3 & \begin{tabular}[c]{@{}l@{}}Camera\\ (DCS700L)\end{tabular} & D-Link & 384 \\ \hline
4 & \begin{tabular}[c]{@{}l@{}}Camera\\ (DCS5030L)\end{tabular} & D-Link & 410 \\ \hline
5 & \begin{tabular}[c]{@{}l@{}}Smart Socket\\ (DSPW215)\end{tabular} & D-Link & 672 \\ \hline
6 & Chromecast & Google & 297 \\ \hline
7 & Home Control & Google & 529 \\ \hline
8 & Smart Socket & Oittm & 394 \\ \hline
9 & Hue Light & Phillips & 644 \\ \hline
10 & Smart Things & Samsung & 587 \\ \hline
11 & \begin{tabular}[c]{@{}l@{}}Smart Bulb\\ (LB100)\end{tabular} & TP-Link & 482 \\ \hline
12 & \begin{tabular}[c]{@{}l@{}}Camera\\ (NCS250)\end{tabular} & TP-Link & 587 \\ \hline
13 & \begin{tabular}[c]{@{}l@{}}Camera\\ (NCS450)\end{tabular} & TP-Link & 494 \\ \hline
14 & \begin{tabular}[c]{@{}l@{}}Smart Socket\\ (HS100)\end{tabular} & TP-Link & 452 \\ \hline
15 & \begin{tabular}[c]{@{}l@{}}Smart Socket\\ (HS110)\end{tabular} & TP-Link & 387 \\ \hline
\end{tabular}%
%}
\end{table}

\subsection{Feature costs}
\label{subsec:feature-costs}

As described earlier, the extraction of each feature has a cost. We argue that the cost of feature extraction is a function of three factors: the computational power (or computational complexity) of the extraction process, the memory used, and the confidentiality of the information related to that feature. Therefore, we break the cost into three components, each corresponding to compute power, memory and privacy.
Finding the exact cost due to each component and how to integrate those component costs into a single cost value is outside the scope of this work. Instead, we simplify the cost of each component to be in one of three levels \{\texttt{low, medium, high}\}. We illustrate the concept of assigning costs to feature using examples in Appendix~A.
%To elaborate, if a register (hash table) is required to compute a feature, we assign a cost of \texttt{medium} (\texttt{high}).
%In the case of memory, if no additional memory (other than that used for storing a packet header) is used to extract a feature, we assign a cost of \texttt{low} to that feature.
%We define the cost matrix $\mathbf{c}$ to indicate the cost associated with each of the 70 extracted features. The feature set is divided based on three different criteria, on which the loss is modeled. Firstly, we analyze the memory of computation required for the extraction of the features from the network traffic. Next, we move to the complexity of the algorithm required for the extraction of the feature and lastly, we look into the fact of how intrusive to privacy the extraction of the said feature is. For example, let us consider the feature $\texttt{'dns\_num\_ans'}$, which is the number of answers returned from the DNS queries. The memory required for the computation of the feature is done by a single register since it is only a counting problem. So, we define the memory cost as \texttt{Medium}. Next, we see that for the extraction of the feature, we had no calculations to be performed like calculating the mean or searching or sorting. So, we assign a \texttt{Low} complexity cost to the feature. Finally, since the feature is extracted from the DNS query, it is considered a higher level of an intrusion than just extracting from the packet header. So, we assign a \texttt{High} privacy cost of to the feature. 
The total cost of a feature $f_i$ can be defined as:
\[\text{Cost}_i = g_i(\text{compute cost}, \text{memory cost}, \text{privacy cost}), \]
where $g_i(.)$ is a function computing the total cost of the feature extraction process. 
For our purposes, we define $g_i, 1 \le i \le m$ to be the median of the three input cost components. While there are other ways of integrating component costs (like, ``cost is always \texttt{high} if privacy cost is \texttt{high}''), we stick to this simple definition for our work here. For example, consider the feature of `{\em connection length, in  number of packets}'. To extract this feature, connection (that is, a 5-tuple flow) identifier has to be hashed and stored in a data structure such as a hash table, and the number of packets needs to be counted~\cite{REX-2017}. 
The computational cost is \texttt{medium} since a hashing is required for every arriving packet, and hash table operations would at worst case be be linear with the number of packets (e.g., for insertion of new 5-tuple flow in the traditional hash table), besides flows have to be regularly removed from the table once they become inactive. 
The memory requirement is assigned as \texttt{high} due to the necessity to maintain a hash table. 
The privacy cost is considered \texttt{low}, as only packet counts of connections are extracted, and no private information (e.g., visited websites) is extracted. Therefore, the cost of extracting {\em connection length} is \texttt{medium}. 
For the experiments in this work, we assign the following values to the above discretizations of the cost values: \{Cost(\texttt{low)=1}, Cost(\texttt{medium)=2}, Cost(\texttt{high)=3}\}. However, note that these values are only for illustrative purposes and any other set of cost values can be considered.

\subsection{Loss matrix, $\mathcal L$}
\label{subsec:loss-eval}

We introduced the concept of misclassificaiton loss in Section~\ref{subsec:loss}. Ideally, the loss matrix is an input provided by the users, based on how they perceive the loss due to misclassification errors. For the purpose of this work, we define the loss matrix as follows.

When the classifier classifies a device $d_i = <t_i, b_i>$ as $d_j = <t_j, b_j>$, the loss value $l_{i,j}$ is calculated as $l_{i,j} = 2*f(t_i, t_j) + f(b_i, b_j),$ where $f(x_i~, x_j) = 1$ if $x_i \neq x_j$ and $0$ if $x_i = x_j$.
Taking every device pair, we follow this definition to derive the values for the loss matrix $\mathcal L$ in the experiments here. It is to be noted that we are giving a higher value of loss if the classifier cannot classify the type of the device even if it gets the right brand. For example, if the classifier classifies a Samsung camera as a Sony camera, the loss value $l_{i,j} = 2 \times 0 + 1 = 1.$ But if it classifies the device as Samsung (Smart things) Hub, then the loss $l_{i,j} = 2*1 + 0 = 2$.

\vspace{-0.2cm}
\subsection{Comparison of classifiers}
\label{subsec:compare-classifiers}

The cyber risk score $R(\mathbf{v})$ of a feature vector $\mathbf{v}$ depends on the misclassification matrix which is in turn derived from the confusion matrix of the classifier. As such, the choice of classifier plays an important role in evaluating the performance of the system. %From the range of available classifiers, we have selected a few within the scope of the paper such that each of the classifiers depicts a unique type of classification mechanism and so, compared them with respect to their $\text{F}_1$ accuracy scores and their run-time complexities. 
We selected four different classification models---Gaussian Naive-Bayes classifier, SVM (support vector machine) with RBF kernel, Decision tree and Random Forest, and compared their performances based on accuracy and the execution time. 
We use $\text{F}_1$-score (the harmonic mean of precision and recall) to represent the accuracy of the classifier. %For the support vector classifier, we use a RBF kernel with $\gamma$, the parameter for non-linear hyperplanes, set to $\frac{1}{m\times \sigma_{X}}$, where $X$ is the input training dataset.
%Since the Naive-Bayes classifier is linearly dependent only on $m$ (the number of features) and the number of training records $N_{\texttt{train}}$, it has the least complexity among the four selected. Following that is the Decision tree classifier, which has a complexity of $\mathcal{O}( m \times N_{\texttt{train}} \log(N_{\texttt{train}}) )$. The computational complexity of Random Forest classifier is linearly dependent on $m$, $N_{\texttt{train}}$, the number of variables to be sampled, the maximum depth of each tree and the number of trees to be built. For SVM, the computation complexity is at least quadratic in the number of training samples $N_{\texttt{train}}$.

\begin{figure}
\centering
\includegraphics[width=0.9\linewidth]{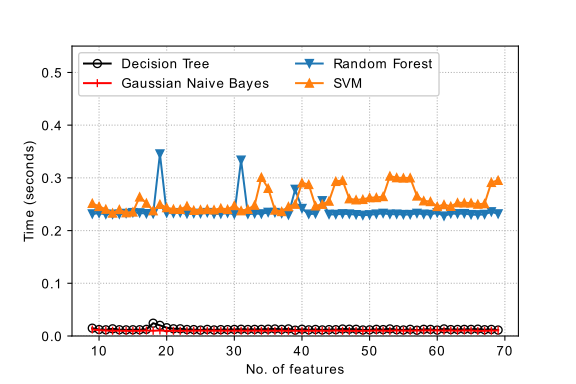}
\caption{Execution time of classifiers for varying number of features $m$}
\label{fig_time_class}
\vspace{-0.2cm}
\end{figure}

All the experiments in this work are carried out on an 8-core Intel Core i7-2600 running at 3.8GHz CPU and equipped with 16GB of RAM.
For comparing the time of execution, we consider both the training and the testing times of the classifiers. Fig.~\ref{fig_time_class} plots the time taken for each classifier for training and predicting on the traffic session of IoT devices. The X-axis here (as well as in all the figures in this paper) denotes the number of features considered for an experiment; therefore, each point on the X-axis corresponds to an independent run of the experiment with that many number of features given as input. Since the number of experiments increases with the number of features, we limit the total number of features provided as input in each experiment to 69; beyond 69 features, the experiments did not provide any new insights.
%We used Random Forest with 100 decision trees, with maximum depth of each tree being~5.
From Fig.~\ref{fig_time_class}, we observe that the Decision tree classifier and the Naive-Bayes classifier take the least time for model building and prediction. 
%gives the best performance while the random forest and the SVM performs the worst among the four. %\todo{(i) SVM doesn't look worst};% (ii) update legend to "SVM"}

\begin{comment}
\begin{figure*}
\begin{minipage}{0.32\textwidth}
\centering
\includegraphics[width=1.1\linewidth]{time_cc_new_100.png}
\caption{Execution time of classifiers for varying number of features $m$}
\label{fig_time_class}
\end{minipage}\hspace{0.2cm}
\begin{minipage}{0.32\textwidth}
\centering
\includegraphics[width=1.1\linewidth]{Figures/accuracy_cc_new_100.png}
\caption{Classification accuracy as a function of the number of features $m$}
\label{fig_acc_com}
\end{minipage}\hspace{0.2cm}
\begin{minipage}{0.32\textwidth}
\centering
\includegraphics[width=1.1\linewidth]{Figures/timet_updated.png}
\caption{Execution time; $\lambda \rightarrow \infty$}
\label{fig:sim_time}
\end{minipage}
\end{figure*}
\end{comment}

%\subsubsection{Accuracy}
%The $\text{F}_1$-score of the different classifiers are plotted in Fig.~\ref{fig_acc_com}. Observe that the accuracy for all the classifiers are very high and close to each other.

\begin{figure}[t]
\centering
\includegraphics[width=0.9\linewidth]{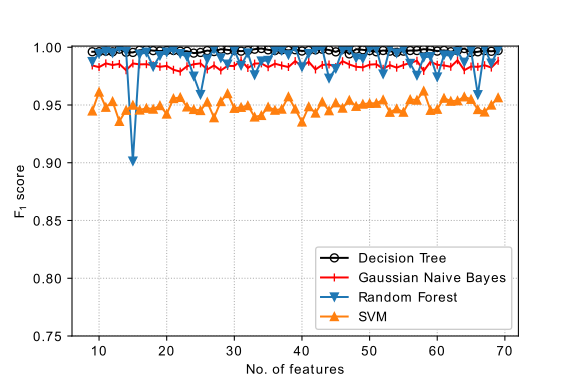}
\caption{Classification accuracy as a function of the number of features $m$}
\label{fig_acc_com}
\vspace{-0.2cm}
\end{figure}

We also plot the accuracy of the different classifiers in Fig.~\ref{fig_acc_com}; all of them achieve high $\text{F}_1$-scores. Since Decision tree classifier has much less time of execution, while also having high $\text{F}_1$-score, henceforth we use Decision tree for the rest of the experiments to evaluate cost-aware feature selection algorithms. 

\subsection{Evaluation of algorithms without budget constraint}
\label{subsec:eval-no-budget}

Next, we compare the five different algorithms discussed in sections \ref{sec:cross} and \ref{sec:others}, by varying the number of features $m$ and with no budget constraint, $\lambda \rightarrow \infty$. We evaluate their performance based on (i) the time of execution and (ii) the cyber risk score.
\par 
%We aim to show the variation of the performance of the algorithms with the increase in the number of features which is independent of the exact features. Thus, we randomly set an order for the feature set $\mathbf{f}$ and repeat the experiment for the different ordering of the feature vector $\mathbf{f}$ and finally take the average of the outputs. 
%Each feature $f$ has a distribution and thus, can be represented using a set of population parameters like the mean, median, variance, etc. 
%In this case, we use the average of the mean, median, variance, and the distribution similarity to calculate the feature score of each feature $f_k$.
As mentioned above, for each experiment, we need to provide a set of features; and the algorithms are supposed to find the best features from the given input set of features. To do this, for each evaluation, we can start with a minimum feature set and increase the set by one feature for each experiment. That is, since we have 69 features, if we start with a singleton set and keep adding one feature for every new experiment, we would need 69 experiments. However, this is under the assumption that, we know the sequence in which the features should be added to the set. Consider adding the most discriminatory feature as the last one to the growing set of features; this means no other experiment would use this important feature for evaluation. On the other hand, the ideal case would be to generate all possible sequences, 69! in our case, and carry out experiments by expanding set from a minimum size to the maximum of 69 features, for each sequence.
%For each experiment, we need to provide as input, a sequence of features, and the algorithms are supposed to find the best features from the given input sequence. 
%As we are considering 69 features for IoT device classification, there are 69! sequences we can provide as input for evaluation purposes. 
Since this is not computationally feasible, unless stated otherwise, we generate sequence of {\em ranked} features. 
%vector $\mathbf f$: (i)~ranked features, and (ii)~random feature ordering. 
To this end, we use the feature ranking mechanism described in our previous work~\cite{desai2019feature} where the features are ranked according to their ability to discriminate between every pair of devices by combining the results from multiple statistical tests. %Observe that, providing rank features is supposed to be useful for two of the greedy algorithms---RGA and VGA. 
%For random feature ordering, we randomly generate a sequence of features. 

%Tell why we are using feature ranking and what is the feature ranking

In order to compare and evaluate the algorithms, we take the full set of ordered feature vector  $\mathbf{f}$ (obtained via ranking),
%or random selection), 
and select the first nine features from it as $\mathbf{\bar{f}} = \left[f_1, f_2, \dots , f_9 \right]$. For each subsequent iteration, we add the next feature in the ranked order to our selected feature vector and calculate the time of execution and the risk score of all the five algorithms using the new set of selected features.
Thus for the $t^{th}$ iteration, the selected feature vector is represented as $\mathbf{\bar{f}} = \left[f_1, f_2, \dots, f_{t+9} \right]; t = 0, 1, 2 , \dots, 61$, where $|\mathbf{\bar{f}}| = t+9$. Therefore, the algorithm must be iterated 61 times, such that, when $t= 61$, all the 69 features will be explored.
We proceed to the analyses next.

%However, it is to be noted that with the changing orders of the features in the feature set, the risk scores calculated for the selected features in each of the iterations may differ, the overall pattern of the graph will remain the same. And since we aim to show the variation of the performance of the algorithms with the increase in the number of features which is independent of the exact features, the order of the features in $\mathbf{f}$ does not play an important role in our analysis.

\subsubsection{Execution time analysis}
The comparison of the execution times of the different algorithms help us to get an estimate of their computational complexities in practice. 
%It also gives us an understanding of the performance of the algorithms on devices with limited computational resources. 
%(\ref{sec:cross}, \ref{sec:others}) For example, an algorithm which is computationally more intensive, will be more difficult to run on a device with limited computational power, than an algorithm with lower complexity.
In this section, we compare the time of execution for the CE algorithm with respect to the brute force and the greedy algorithms, under no budget constraint ($\lambda \rightarrow \infty$).

%experiments, we have used a Intel Core i7-2600 with 8 cores with 2 threads each and 3.8GHz CPU with AMD CEDAR Radeon HD 8350 Series GPU with 1GB of VTT memory and 512MB of VRAM memory.It is also equipped with 16GB of DDR3 RAM running on Ubuntu 18.04 Bionic.

\begin{figure}
\centering
\includegraphics[width = \linewidth]{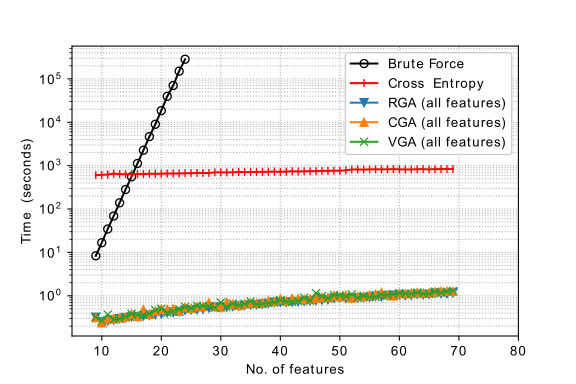}
\caption{Execution time as a function of number of features $m$ with $\lambda \rightarrow \infty$}
\label{fig:sim_time}
\vspace{-0.3cm}
\end{figure}

From Fig. \ref{fig:sim_time}, we observe that the time of execution for the brute force algorithm increases exponentially as the number of features in the feature set increases. Thus, though the brute force algorithm has a low time of execution for small feature sets, it is extremely high for larger ones. This is why we limit the experiment on the brute force to 25 features. The greedy algorithms perform the best with respect to the execution time, which gradually increases with increasing $m$. Though the proposed CE algorithm takes more time than the greedy algorithm, its execution time is much lower than the brute force approach.  Also, the growth in run-time with increasing number of features is marginal for the CE algorithm, thereby indicating the ability of the algorithm to scale.

%We expect similar trend for executions of the algorithms under budget constraint,  because the time of execution of an algorithm depends on $m$ and the role of the budget $\lambda$ is only to filter the features from $\mathbf{\bar{f}}$ to select the optimum feasible feature set.

\subsubsection{Risk analysis}
\begin{figure}
\centering
\includegraphics[width = \linewidth]{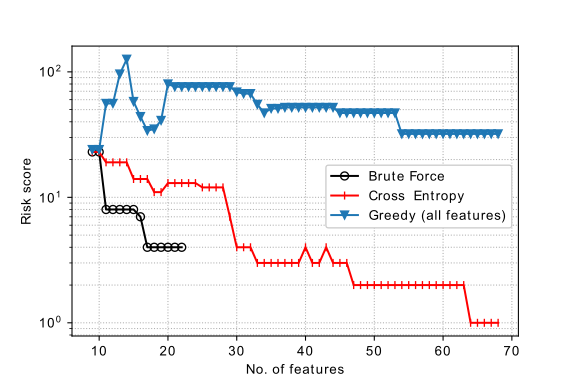}
\caption{Risk score $R(\mathbf{v})$ as a function of number of features $m$ with $\lambda \rightarrow \infty$}
\label{fig:sim_loss}
\vspace{-0.4cm}
\end{figure}

The risk score $R(\mathbf{v})$ is another important parameter to evaluate the performance of the feature selection algorithms as we aim to minimize the risk score by the optimal selection of features. 
The graphs for the risk scores of the five different algorithms under consideration without any budget constraint is plotted in Fig. \ref{fig:sim_loss}.
Since this scenario does not constrain the budget, i.e., $\lambda \rightarrow \infty$, note that,  the three greedy algorithms will always choose all the features provided and thus, no difference is expected in the performance (in terms of accuracy and risk score) of the three different greedy approaches. 
We observe that the brute force algorithm gives the best risk score since it performs an exhaustive search over all possible combinations of the provided features. However, it can also be observed that the cross entropy algorithm is close in performance to that of the brute force method, and ultimately converges to the minimal risk score of~1.
Due to this convergence, it may be concluded that any feature more than the first~63 (in the ranked order) is redundant in the classification of the devices when using the cross-entropy algorithm. 
In addition to this, it can also be observed that though the greedy algorithms select all the features, they do not give the best results. This highlights the necessity for the selection of an optimal set of features, and that the selection of a higher number of features does not necessarily imply a better risk score. 
%We highlight here that, though the features provided here were ranked as per~\cite{desai2019feature}, the ranking as such does not consider risk of misclassification. 
%{\color{blue} We see from Fig. \ref{fig:sim_loss}, the risk score for the cross entropy method decreases sharply when the $29^{th}$ feature, \texttt{ssdp duration} feature is added. For the greedy algorithm, there is a peak when we add the $11^{th}$ and $12^{th}$ features \texttt{mqtt pkt no, mqtt time mean} respectively and it again drops when the $13^{th}$ feature \texttt{mqtt time min}.}

 %Infact, the greedy algorithms show a poorer performance when using more than 63 features due to possible over-fitting.
%\par Next, we observe that among the greedy algorithms, the algorithm selects the features according to the cyber risk score which was calculated taking 1 feature at a time. Following it, the greedy algorithm based on the cost of each feature performs better, and the algorithm with respect to the density performs the best among the three. 

\subsection{Evaluation of algorithms under budget constraint}
\label{subsec:eval-budget}

%\subsubsection{Cross Entropy}
\begin{figure}
\centering
\includegraphics[width = \linewidth]{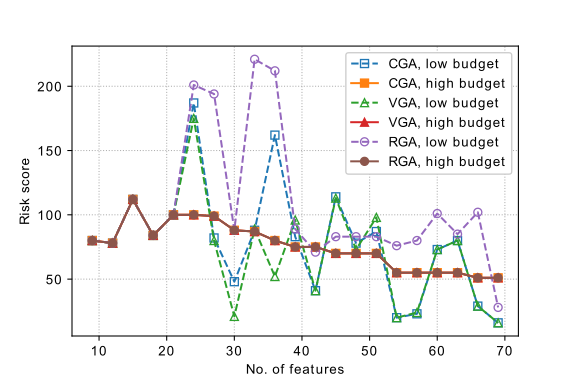}
\caption{Risk score $R(\mathbf{v})$ as a function of the number of features $m$ for greedy algorithms with low and high budget $\lambda = \left\{50,200\right\}$}
\label{fig:greedy_budget}
\vspace{-0.3cm}
\end{figure}
% \begin{figure}
% \centering
% \includegraphics[width = 0.7\linewidth]{Figures/shapley_21.png}
% \includegraphics[width = 0.7\linewidth]{Figures/shapley_24.png}
% \caption{Shapley Values for ranked features for 21 and 24 features repsectively}
% \label{fig:shapley}
% \vspace{-0.3cm}
% \end{figure}
\begin{table}[t]
\centering
\caption{Comparison of Greedy Algorithms}
\label{tab:comp_greedy}
\begin{tabular}{|c|c|c|c|c|c|c|}
\hline
 & \multicolumn{3}{c|}{\textbf{\begin{tabular}[c]{@{}c@{}}Feature Budget \\ $\lambda = 50$\end{tabular}}} & \multicolumn{3}{c|}{\textbf{\begin{tabular}[c]{@{}c@{}}Feature Budget\\ $\lambda = 200$\end{tabular}}} \\ \hline
 & \textbf{CGA} & \textbf{VGA} & \textbf{RGA} & \textbf{CGA} & \textbf{VGA} & \textbf{RGA} \\ \hline
\textbf{Mean} & 68.28 & 64.82 & 98.03 & 69.79 & 70.04 & 68.11 \\ \hline
\textbf{\begin{tabular}[c]{@{}c@{}}Standard \\ Deviation\end{tabular}} & 41.97 & 36.72 & 51.62 & 18.25 & 18.84 & 20.79 \\ \hline
\end{tabular}
\end{table}

\textbf{Comparison of greedy algorithms:}
We first compare the performance of the three different greedy algorithms under budget constraint to determine the best-performing greedy algorithm. We consider two cases: one with low budget for features, $\lambda = 50$, and another with high feature budget, $\lambda = 200$. We plot the  risk scores of the three greedy algorithms Fig.~\ref{fig:greedy_budget}. 

Similar to previous experiments, each point on the X-axis on Fig.~\ref{fig:greedy_budget} denotes the number of features given as input to the algorithms. However, to limit the number of experiments, we start with nine features and add three features with every new experiment. Therefore the point 30 on the X-axis means, 30 features were given as input to the algorithms in the experiment, and the next point corresponds to 33 features given as input. We observe that, all the three greedy algorithms have similar performance when the budget is high. With high budget, all features given at input can be selected for building the classifier. However, this is not the case for lower budget---under tighter budget constraint, we see that the three algorithms have different performances as they choose different (subsets of) features from the given list of input features. For easier comparison, we also provide the mean and standard deviation of risk scores for each of the algorithms in Table~\ref{tab:comp_greedy}. From these results, it can be seen than VGA performs better than the other two under low budget (while all three have comparable performance under high budget). 

An interesting aspect to note in the low-budget case  is that, the risk scores do not decrease monotonically with increasing features provided at the input, instead we see spikes in the risk score (Fig.~\ref{fig:greedy_budget}). To understand this, let us consider the first spike --- the point on the X-axis where the first 24 ranked features are added. Observe, all three algorithms experience a spike in the risk score at this point. We present the analysis of VGA. Compared to the previous point on the X-axis, i.e., the experiment corresponding to the first 21 features, the features \texttt{mdns\_len mean}, \texttt{mdns\_duration}, \texttt{mdns\_num ans}, and \texttt{http\_time\_mean} were newly added to the 
%search space of the 
classifier model by VGA, but more importantly the features \texttt{dns\_num\_qns} , \texttt{dns\_qry\_cls},  \texttt{http\_len\_mean}, and \texttt{tcp\_keep\_alive} were removed by the algorithm.
In general, a specific characteristic of the greedy approaches that lead to the spikes and valleys is the following: with a limited feature budget, a greedy algorithm selects the maximum number of features possible with the given budget, which might not include the features selected in the previous iteration. In other words, this might lead to the rejection of features with greater discriminating power, thus resulting in the valley (or vice-versa, leading to a spike due to the selection of much better features in the next iteration). Furthermore, in order to get an analytical evaluation of these features, we used the SHAP method~\cite{SAHP-NIPS-2017} to estimate the contributions of different features in classifying the devices. SHAP  uses the concept of Shapley values from Game theory, to explain the predictions of a machine learning model in a similar way as computing the contributions of players in a game. Using the SHAP method, we note that, the second-most important feature in the list of first 21 features was \texttt{tcp\_keep\_alive}, and this was removed by VGA in the next experiment when three more features were added. Besides, none of the MDNS related features that were newly added by VGA were in the top eight most contributing features. We observe a similar removal of important features in both RGA and CGA, thus rendering them ineffective in selecting the most discriminative features to minimize risk.

Based on the above experiments, to compare the performance of the greedy methods with the CE algorithm, we select VGA as the candidate as it gives the best performance among three greedy approaches. \\

\begin{figure}
\centering
\includegraphics[width = \linewidth]{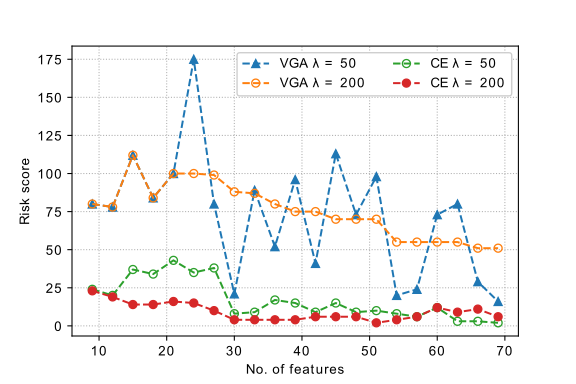}
\caption{Risk score $R(\mathbf{v})$ as a function of the number of features $m$ with discrete $\lambda$ for CE and VGA algorithms for \textbf{ranked features}}
\label{fig:cebudget}
\vspace{-0.3cm}
\end{figure}

\noindent \textbf{Comparison of CE algorithm and VGA, for ranked and random ordering of features:} The risk score calculated is dependent on the ordering of the features $\mathbf{f}$ and hence to test the performance of the algorithms, we carry out the experiments for not only the ranked feature vector, but also for randomly ordered features. The risk score $R(\mathbf{v})$ is a function of the feature budget $\lambda$. Therefore, similar to the previous scenario, we compare the risk score as a function of the number of features $m$ for a low budget ($\lambda = 50$) and high budget ($\lambda = 200$). For CE, we set the number of samples, $\eta = 1000$ and $T_{\max} = 500$; and compare CE with and VGA algorithms. Fig.~\ref{fig:cebudget} plots the results when the input provided was the ranked feature vector. CE is seen to outperform VGA under both low and high budgets. The CE algorithm consistently achieves lower risk score with increasing features, in comparison to VGA.

%It is also worthwhile to note that the risk based greedy algorithm with lower budget performs better than the greedy algorithm with higher budget. This is due to the fact the greedy algorithm chooses all the features which can be got within the feature budget. Thus, for a larger feature budget, the risk based greedy method is using a larger number of features, which not necessarily gives the best classification result and hence the inferior risk scores. This further highlights our point that simply using the maximum number of features using greedy is not an optimal way for feature selection.\\

\begin{figure}
\centering
\includegraphics[width = \linewidth]{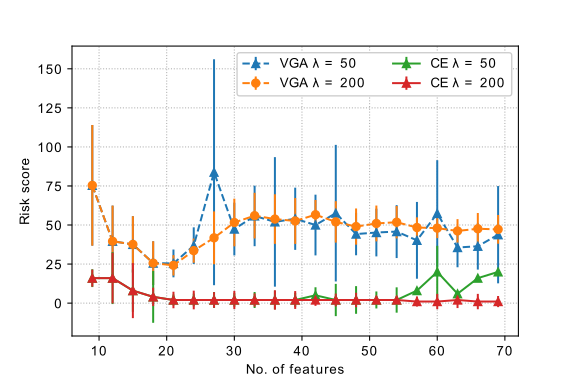}
\caption{Risk score $R(\mathbf{v})$ as a function of the number of features $m$ with discrete $\lambda$ for CE and VGA algorithms for \textbf{randomly ordered features}}
\label{fig:cebudget_rand}
\vspace{-0.3cm}
\end{figure}

Under low budget constraint, we can clearly see that the CE algorithm outperforms the greedy method for all feature lengths $m < 70$. %As we will see later, the CE algorithm requires more number of samples and higher iterations to converge to a lower risk score, when the number of features is high\todo{revise based final figure}. 
We observe a similar trend in Fig.~\ref{fig:cebudget_rand}, when randomly ordered features are provided as input. In this scenario, we evaluate the results for five different randomly ordered feature sequences, and show the standard deviation of the risk scores obtained. We observe that the risk score's standard deviation is high for small feature length, which is quite intuitive as the small number of features (given as input) can be significantly different with every random sample. Thus the risk score also varies with different input features provided. We also note that the standard deviation  of the risk scores of the CE algorithm is always low (and extremely low for the the high-budget scenario), showing the utility of the proposed algorithm.

%We observe from the figure that the CE algorithm gives the highest risk for low $\lambda$ and a large $m$. Since it is difficult to deduce a definite conclusion from the figure, we plot the graphs for discrete $\lambda = 10, 50, 100, 150, 200$ for varying $m$.

\noindent \textbf{Comparison of CE algorithm and VGA under varying budget:}
Finally, we compare the risk scores for the CE algorithm for varying feature budgets. The features are provided in ranked order.
Since we now have two parameters to vary ($m$ and $\lambda$), for this set of experiments, we keep the number of samples $\eta = 250$ and the number of iterations $T_{\max} = 50$ for the CE algorithm.
%Fig.~\ref{fig:cem_budget} provide a three-dimensional analysis, representing the risk score as a function of, both, the budget constraint ($\lambda$) and the number of features $(m)$.
%We observe that the risk score obtained under a low budget increases as we increase the number of features. However, if we use a higher budget, then the risk score remains low even for a larger number of features. The CE algorithm gives high risk for large values of $m$ under low budget, because the algorithm rejects the initial samples whose total cost is greater than the budget. Hence, for large $m$ and a small $\lambda$, the algorithm rejects most of the initial samples, prohibiting the algorithm to select the optimal set of features possible, thus resulting in a high risk score. However, on increasing the budget, the risk score is seen to decrease. 
%decreases to a large extent. To get an overall picture of the variation of risk score with changing $\lambda$ and $m$, we plot Fig. \ref{fig:cem_budget}.
We also conducted a similar experiment for VGA.  In order to have a visual comparative analysis of the performance of the CE algorithm with the greedy algorithm, we take the difference between their risk scores for each $m$ and $\lambda$. The resultant graph for $R_{\texttt{CE}}(\mathbf{v}) - R_{\texttt{VGA}}(\mathbf{v})$ is plotted in Fig.~\ref{fig:diff_budget3d_in2d} (see Appendix B for the 3D version of the graph). The part of the graph which is greater than 0 (above the $z=0$ plane) denotes the conditions where the cross-entropy method has greater risk than the greedy algorithm. %
The risk score achieved by CE is lower than that of VGA for most cases, except when both, the budget is very low, and the number of features is high. 
For large $m$ and a small $\lambda$, the CE algorithm rejects most of the initial samples, prohibiting the algorithm to select the optimal set of features possible, thus resulting in a high risk score. However, on increasing the budget, the risk score is seen to decrease.
Besides, as seen in the previous set of experiments, the performance of CE can be improved by increasing the number of samples and the number of iterations. 
\begin{comment}

\begin{figure}
\centering
\includegraphics[width = \linewidth]{Figures/CEM_budget_new_to235.png}
\caption{Risk score $R(\mathbf{v})$ as a function of the number of features $m$  and the feature budget $\lambda$ for cross entropy algorithm}
\label{fig:cem_budget}
\end{figure}
\end{comment}

\begin{figure}
\centering
\includegraphics[width=0.95\linewidth]{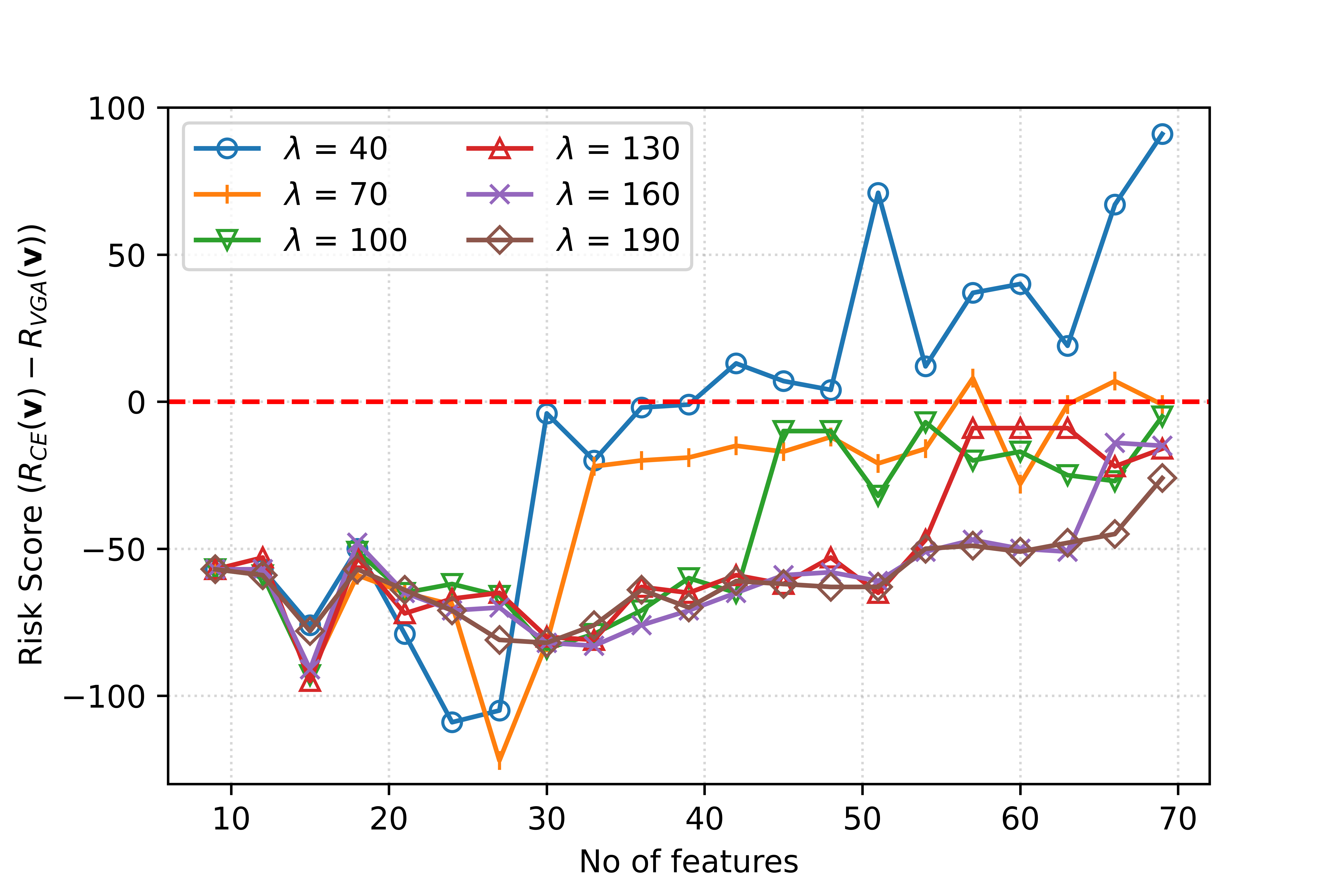}
\caption{$R_{\texttt{CE}}(\mathbf{v}) - R_{\texttt{VGA}}(\mathbf{v})$ as a function of the number of features $m$  and the feature budget $\lambda$. The features are in ranked order.}
\label{fig:diff_budget3d_in2d}
\vspace{-0.3cm}
\end{figure}

%% file: tex/conclusions.tex
\section{Conclusions} \label{Sec:conclusions}

In this work, we motivated, defined and focused on the problem of feature selection for IoT device classification considering the fact that there is a budget constraint for features in practice. For this purpose, we also defined the notion of risk of misclassification. To obtain the optimal solution to this problem, one has to perform a combinatorial search over the solution space. Therefore, we developed a cross entropy based algorithm for solving the optimization problem. We carried out experiments using traffic of real IoT devices; our experiments showed that not only is CE algorithm practical and much faster than the brute force approach, it also obtains low risk score for classification and performs better than than value-based greedy algorithm in most cases.

%% file: tex/appendix.tex
\section*{Appendix}

\subsection{Feature cost}

In this section we demonstrate how feature costs are estimated for a given IoT device.
% Finding the exact cost due to each component and how to integrate those component costs into a single cost value is outside the scope of this work. Instead, we simplify the cost of each component to be in one of three levels \{\texttt{low, medium, high}\}. To elaborate, if a register (hash table) is required to compute a feature, we assign a cost of \texttt{medium} (\texttt{high}).
% In the case of memory, if no additional memory (other than that used for storing a packet header) is used to extract a feature, we assign a cost of \texttt{low} to that feature.
We defined the cost vector $\mathbf{c}$ to indicate the cost associated with each of the  extracted features. The feature cost is divided into three different components. First, we analyze the memory required for the extraction of the features from the network traffic. Next, we focus on the computational complexity of the algorithm required for extracting the feature. And lastly, we analyze the intrusiveness of a feature into the privacy aspects of a device or user. 
For example, consider classification of three IoT devices. Assume, 
\texttt{dns\_num\_ans} and
\texttt{tcp\_keep\_alive} are the two most useful features with equal discriminative power, such that only one of these two features is sufficient for classifying the three devices accurately. 
Let us consider the feature \texttt{dns\_num\_ans}, which denotes the the number of answers returned for DNS queries. While only a register is required to store the (running) value of this feature, a small buffer has to be maintained to temporarily store (unprocessed) DNS answers extracted from packet payloads (to absorb traffic bursts); therefore, we define the memory cost as \texttt{medium}. Similarly, this feature is computed by simple additions to the counter in the register. Therefore we can assume a \texttt{low} computational complexity. Finally, since the feature is extracted from the DNS responses (i.e., packet payload), which can leak considerable information (e.g., see~\cite{DNS-ads-profit-2011}), we assign a \texttt{high} privacy cost to the feature. Now consider the feature \texttt{tcp\_keep\_alive}---the number of TCP keepalive packets. The cost of memory, compute power and privacy, are all \texttt{low} for this feature (only a counter needs to be maintained to extract feature values from packet headers). Therefore, between the two features, \texttt{dns\_num\_ans} and
\texttt{tcp\_keep\_alive}, the latter is a better feature when we consider cost, given both are equally good at classifying the three devices.
% Therefore, the total cost of a feature $f_i$ can be defined as:
% \[\text{Cost}_i = g_i(\text{compute cost}, \text{memory cost}, \text{privacy cost}), \]
% where $g_i(.)$ is a function describing the total cost of the feature extraction process based on the three components of compute power, memory and privacy. 
% We define $g_i, 1 \le i \le m$ to be the median of the three input cost components. While there are other ways of integrating component costs (like, ``cost is always \texttt{high} if privacy cost is \texttt{high}''), we stick to this simple definition for our work here. For example, consider the feature of `{\em connection length, in the number of packets}'. To extract this feature, connection (that is, a 5-tuple flow) identifier has to be hashed and stored in a data structure such as a hash table, and the number of packets needs to be counted~\cite{REX-2017}. 
% The computational cost is \texttt{medium} since a hashing is required for every arriving packet, and hash table operations can be linear in worst cases (e.g., for insertion of new 5-tuple flow in the traditional hash table), besides flows have to be regularly removed from the table once they become inactive. 
% The memory requirement is assigned as \texttt{high} due to the necessity to maintain a hash table. 
% The privacy cost is considered \texttt{low}, as only packet counts of connections are extracted, and no private information (e.g., visited websites) is extracted. Therefore, the cost of extracting {\em connection length} is \texttt{medium}. %\todo{right? --yes}.

Table~\ref{tab:feature_costs_details} provides the logic we have used to define the three cost components for extracting any given feature from network traffic. 
%We calculate the cost of each feature similarly and obtain the feature cost vector given by $\mathbf{c}$. 
%The costs for different memory, complexity and privacy tasks are calculated according to the values in Table \ref{tab:feature_costs}.
%For example, if the memory cost is less than 40 bytes, we assign a memory cost of \texttt{Low} to it. Similarly, for the privacy cost, if we are extracting the features from the DNS or Session information, then we associate a \texttt{High} cost to it. Thus, we get a memory, complexity and privacy cost for each of the 93 features. 
To define the total cost of extracting a feature, we take the median of all three costs (memory, compute and privacy). 
%Since we are using a linear model to combine the feature costs, we define the total cost of extraction of the feature as the sum of the three costs. For example, if the memory cost of a said feature is \texttt{Low}, the complexity cost is \texttt{Medium} and the privacy cost is \texttt{High}, then we calculate the total cost of the feature using the median of the three, given as $\textbf{median}(\texttt{Low}, \texttt{Medium},\texttt{High}) =\texttt{Medium}.$

\begin{table}[h]
\centering
\caption{Cost logic considered for experiments}
\resizebox{\columnwidth}{!}{%
\begin{tabular}{|c|c|c|c|c|}
\hline
\multirow{2}{*}{\textbf{\#}} & \multicolumn{3}{c|}{\textbf{Parameters}} & \multirow{2}{*}{\textbf{Cost}} \\ \cline{2-4}
 & memory & compute power & privacy &  \\ \hline
1 & \begin{tabular}[c]{@{}c@{}}Single register \\ (constant memory)\end{tabular} & \begin{tabular}[c]{@{}c@{}}Counters \\ (e.g., $\min, \max$, etc.) \end{tabular} & \begin{tabular}[c]{@{}c@{}}Features extracted \\ from packet headers\end{tabular} & \texttt{low} \\ \hline
2 & \begin{tabular}[c]{@{}c@{}}Small buffer \\ (e.g., queue) \end{tabular} & \begin{tabular}[c]{@{}c@{}}Maintenance of \\ hash tables \end{tabular} & \begin{tabular}[c]{@{}c@{}}Extraction of \\ application data (e.g., URL)\end{tabular} & \texttt{medium} \\ \hline
3 & \begin{tabular}[c]{@{}c@{}}Hash table or \\ multiple registers\end{tabular} & \begin{tabular}[c]{@{}c@{}}Pattern matching \\ or sorting \end{tabular} & \begin{tabular}[c]{@{}c@{}}Features extracted from \\ packet payloads \end{tabular} & \texttt{high} \\ \hline
\end{tabular}%
}
\label{tab:feature_costs_details}
\end{table}

% \begin{figure}[h]
% \centering
% \includegraphics[width=0.95\linewidth]{FinalFigures/diff_budget_new_to235_plane.png}
% \caption{$R_{\texttt{CE}}(\mathbf{v}) - R_{\texttt{VGA}}(\mathbf{v})$ as a function of the number of features $m$  and the feature budget $\lambda$}
% \label{fig:diff_budget3d}
% \vspace{-0.3cm}
% \end{figure}

\begin{figure}[h]
\centering
\includegraphics[width=0.95\linewidth]{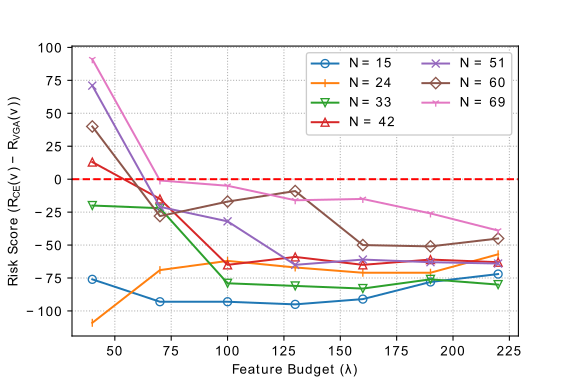}
\caption{$R_{\texttt{CE}}(\mathbf{v}) - R_{\texttt{VGA}}(\mathbf{v})$ as a function of the feature budget $\lambda$}
\label{fig:diff_budget3d}
\vspace{-0.3cm}
\end{figure}

\subsection{Comparison  of  CE  algorithm  and  VGA  under  varying budget}
As described in Section~\ref{subsec:eval-budget}, we are interested in studying how the risk scores of the cross entropy algorithm and the value-based greedy algorithm  change with varying number of features used ($m$) as well as for different feature budgets ($\lambda$). Fig.~\ref{fig:diff_budget3d} presents an alternate 2D graph plotting the difference between the two risk scores, $R_{CE}(\mathbf{v}) - R_{VGA}(\mathbf{v})$, 
as a function of $\lambda$. The area of the figure below the plane $[R_{CE}(\mathbf{v}) - R_{VGA}(\mathbf{v}) = 0]$ denotes the regions where the CE algorithm outperforms VGA.  As discussed in Section~\ref{sec:simulations}, the greedy approach has lower risk only for the cases where the number of features is high and the budget is low; and this is mainly due to the computational constraints that limit our CE algorithm to converge in our simulations.